\documentclass[aos,preprint]{imsart}

\RequirePackage[OT1]{fontenc}
\RequirePackage{amsthm,amsmath}
\RequirePackage[numbers]{natbib}
\RequirePackage[colorlinks,citecolor=blue,urlcolor=blue]{hyperref}

\arxiv{arXiv:0000.0000}

\startlocaldefs
\numberwithin{equation}{section}
\theoremstyle{plain}

\endlocaldefs

\usepackage{array}
 \usepackage{amsmath,amssymb}
 \usepackage{amsthm}
 \usepackage{bm}
 \usepackage{latexsym}
 \usepackage{CJK}
 \usepackage{multirow}
 \usepackage{caption}
 \usepackage{graphicx,color}
 \usepackage{amssymb,amsfonts}
 \usepackage{hyperref}
 \usepackage{times}
 \usepackage{enumerate}
\usepackage{geometry}
\usepackage{float}
\usepackage{endnotes}
\usepackage{mathtools}
\usepackage{amscd}
\usepackage{bbm}
\usepackage{multirow}
\usepackage{booktabs}

\usepackage[dvipsnames]{xcolor}

\newtheorem{theorem}{Theorem}[section]

\newtheorem{lemma}{Lemma}

\DeclareMathOperator*{\argmax}{argmax}

\newcommand{\var}{{\rm Var}}

\begin{document}

\begin{frontmatter}
\title{Community Detection with Dependent Connectivity\thanksref{T1}}
\runtitle{Community Detection with Dependent Connectivity}
\thankstext{T1}{This work is supported by NSF Grants DMS 1613190 and DMS 1821198.}

\begin{aug}
\author{\fnms{YUBAI} \snm{YUAN}\thanksref{m1}}
\and
\author{\fnms{ANNIE} \snm{QU}\thanksref{m1}}



\affiliation{University of Illinois at Urbana-Champaign\thanksmark{m1}} 

%

\end{aug}
\begin{abstract}
In network analysis, within-community members are more likely to be connected than between-community members, which is reflected in that the edges within a community are intercorrelated. However, existing probabilistic models for community detection such as the stochastic block model (SBM) are not designed to capture the dependence among edges. In this paper, we propose a new community detection approach to incorporate intra-community dependence of connectivities through the Bahadur representation. The proposed method does not require specifying the likelihood function, which could be intractable for correlated binary connectivities. In addition, the proposed method allows for heterogeneity among edges between different communities. In theory, we show that incorporating correlation information can achieve a faster convergence rate compared to the independent SBM, and the proposed algorithm has a
lower estimation bias and accelerated convergence compared to the variational EM. Our simulation studies show that the proposed algorithm outperforms the popular variational EM algorithm assuming conditional independence among edges. We also demonstrate the application of the proposed method to agricultural product trading networks from different countries.
\end{abstract}

\begin{keyword}[class=MSC]
\kwd[Primary ]{60K35}
\kwd{60K35}
\kwd[; secondary ]{60K35}
\end{keyword}

\begin{keyword}
\kwd{Bahadur Representation}\kwd{high-order approximation}\kwd{product trading network}\kwd{stochastic block model} \kwd{variational EM}
\end{keyword}
\end{frontmatter}

\section{Introduction}
%


Network data has arisen as one of the most common forms of information collection. This is due to the fact that the scope of study not only focuses on subjects alone, but also on the relationships among subjects. Networks consist of two
components: (1) nodes or vertices corresponding to basic units of a system, and (2) edges representing connections between
nodes. These two main components can have various interpretations under different contexts of application. For example, nodes might be humans in social networks; molecules, genes, or neurons in biology networks, or web pages in information networks. Edges could be friendships, alliances, URLs, or citations. The combination of the nodes and the edges defines
a network,  which can be represented by an adjacency matrix  to reflect  direct  connectivities among nodes.

In this paper, we are interested in identifying community structures, such as community detection of cluster nodes which have more concentrated connectivities in a subnetwork. Identifying communities is essential to provide deep understanding of relationships  among nodes within a community and between communities to address scientific, social and political problems \citep{waskiewicz2012friend, bechtel2005lung, garcia2018applications,warnick2018bayesian,ogburn2017vaccines,
lawrence2006network,montanari2010spread}.
In terms of other applications, community detection plays an important role in decomposing original large-scale network  structures \citep{zanin2008complex, valverde2012link, pinheiro2012community} into several subnetworks with more simplified  structures \cite{bui2014modular}, and facilitates scalable computation for further analyses.

The major community detection methods can be summarized in the following  three categories. One approach is to  search a partition of nodes  which optimizes a global criterion over all possible partitions. The  corresponding   criterion  function measures  the  goodness of fit of a partition such as  the modularity \cite{newman2004finding} or profile likelihood \cite{bickel2009nonparametric} on the observed networks \citep{shi2000normalized, le2016optimization, amini2013pseudo} to capture densely connected communities. However, obtaining a global optimum based on this type of criterion is computationally infeasible. In addition, modularity also suffers from the resolution limit \cite{fortunato2007resolution}, which intrinsically ignores small communities.
 The second approach is the spectral method \citep{rohe2011spectral, donetti2004detecting, arias2011spectral}, which recovers dense connectivities through the eigenvectors of the adjacent matrix of the network. One critical drawback of the spectral method is that it lacks robustness in estimation, especially when networks are sparse or consist of high-degree nodes such as  hubs \cite{krzakala2013spectral}. The third approach is the maximum likelihood method for cluster networks. This includes the popular stochastic block model (SBM) \cite{holland1983stochastic} and its extensions to incorporate the heterogeneity of nodes' degrees \cite{karrer2011stochastic,zhao2012consistency}, and latent distance modeling \citep{handcock2007model, hoff2008modeling} to handle overlapping communities \citep{airoldi2008mixed, ball2011efficient}.

The SBM assumes that membership assignments for each node follow a multinomial distribution. Given the community memberships, edges in the same community are randomly generated from a specified  distribution. The common key assumption for SBM algorithms is that connectivities are conditional independent given the membership of nodes. This assumption simplifies the complexity of the model, and the likelihood function can be explicitly formulated. However, the network data are likely dependent among connectivities, which are also considered in several random network modelings \citep{hoff2018additive, lauritzen2018random, kim2016geometry,cheng2014sparse}. For community detection, the conditional independency assumption typically does not hold in practice and therefore could lead to a misspecified model  
\citep{rahman2017link, anagnostopoulos2008influence, von1994correlation}. For example, friendships within a social community or  functional connectivities in brain networks tend to be highly correlated.

In addition, under conditional independence, the community structure can only be identified based on the marginal mean discrepancy of connectivities between within-communities and across-communities. Specifically, as a fundamental assumption of the independent SBM, the marginal mean discrepancy is required to be greater than a sharp threshold to guarantee community detectability (\citep{massoulie2014community, mossel2015consistency}). However, the marginal mean  discrepancy assumption might not hold, while the correlations among edges could be non-negligible and highly informative in identifying community structures. We show that the proposed method is able to incorporate the correlation information to achieve consistent community detection when the marginal mean discrepancy is insignificant.



More recently, the SBM has been extended to address the heterogeneity feature of within-community for multiple network samples. For example, \citep{stanley2016clustering, paul2018random} apply a fixed-effect model through an independent intercept without incorporating information from other networks. 
Alternatively, a random-effects model is proposed to incorporate heterogeneity \cite{pavlovic2015generalised, zhang2018mixed}, which borrows  information from multiple networks. However, both of these approaches require the specification of a distribution for the random effects. In addition, an EM-type algorithm is implemented to integrate out the random-effects, \citep{pavlovic2015generalised, zhang2018mixed} which could be computationally  expensive when the size of the community or the  network size is large.

In this paper, we propose a novel community detection method to jointly model community structures among multiple networks. The proposed method can simultaneously incorporate the marginal and correlation information to differentiate within-community and between-community connectivities. The key idea is to approximate the joint distribution of correlated within-community connectivities by using a truncated Bahadur representation \cite{bahadur1959representation}. Although the approximate likelihood function is not the true likelihood, it is able to maximize the true community memberships and serves as a tighter lower bound to the true likelihood compared with the independent SBM likelihood. Consequently, we identify communities via maximizing the approximate likelihood function, which also serves as a discriminative function for membership assignments of nodes.
In particular, within-community correlations provide an additional community-concordance measurement to capture high-order discrepancy between within-community and across-community networks, and therefore increase discriminative power to identify communities.

The main advantages and contributions of the proposed method can be summarized as follows. The proposed method  incorporates correlation information among connectivities to achieve more accurate community detection than the variational EM method using marginal information only. The improvement of the proposed community detection method is especially powerful when the marginal information is relatively weak in practice. 
In addition, compared to the existing random-effects model, the proposed method is more flexible in modeling the heterogeneity of communities for multiple networks and heterogeneity of correlations among edges. Furthermore, it does not require a distribution specification among within-community connectivities.

In addition, we establish the consistency of the community estimation for the proposed approximate likelihood under a general within-community edge correlation structure and show that the proposed method achieves a faster convergence rate of membership estimation compared to the independent likelihood. In terms of computational convergence, the proposed algorithm achieves a lower estimation bias and a faster convergence rate compared to the variational EM algorithm at each iteration via incorporating additional correlation information. The theoretical development in this paper is nontrivial, since establishing membership estimation consistency is more challenging under the framework of conditional dependency among edges compared to the existing ones assuming the conditional independent model. Furthermore, we show that the convergence of the variational EM algorithm \cite{mariadassou2010uncovering} is a special case of our method under the conditional independent SBM.

Computationally, we develop a two-step iterative algorithm which is not sensitive to initial values as in the standard variational EM algorithm. In addition, compared to the existing fixed-effects SBM with independent intercepts or the random-effects SBM, the proposed method has lower computational complexity, as it does not involve integration of random effects as in \cite{pavlovic2015generalised}, or  estimating the fixed effects for each network as in \cite{paul2018random}. Simulation studies and a real data application also confirm that the proposed method outperforms the existing variational EM significantly, especially when the marginal information of observed networks is weak. 

This paper is organized as follows: Section 2 introduces the background of the proposed method. Section 3 introduces the proposed method to incorporate correlation information for community detection. Section 4 provides an algorithm and implementation strategies. Section 5 illustrates the theoretical properties of the proposed method. Section 6 demonstrates simulation studies, and Section 7 illustrates an application to   world agricultural products trading data. The last section provides conclusions  and some further discussion.

\section{Background and Notation}
In this section, we provide background and notation of the proposed community detection. The stochastic block model (SBM) \cite{holland1983stochastic} is a form of hierarchical modeling which captures the community structure for networks. Consider $M$ symmetric and unweighted sample networks $\bm{Y} = \{\bm{Y}^m\}_{m=1}^M = \{(Y_{ij}^m)_{N\times N}\}_{m=1}^M$ with $N$ nodes for $K$ communities. Let $\{z_i\}_{i=1}^N$ be the membership for each node and $z_i \in \{1,2,\cdots,K\}$, and denote the membership assignment matrix $\bm{Z} = \{(Z_{iq})_{n\times K}\}\in \{0,1\}^{N\times K}$, where $Z_{iq}= \mathbbm{1}\{z_i = q\}$. Here $\bm{Z}$ has exactly one 1 in each row and at least one 1 in each column for no-null communities. The unknown membership $z_i \in \{1,2,\cdots,K\}$ can be modeled as a latent variable from a multinomial distribution:
$$z_{i} \sim Multinominal(1,\alpha_i),$$
where $i=1,\cdots, N$, $\alpha_{i} = \{\alpha_{i1},\cdots,\alpha_{iK}\}$ and $\sum_{k=1}^K\alpha_{ik} = 1$. Given the membership of nodes, the observed edges between two nodes $\{(Y_{ij}^m)_{n\times n}\}_{m=1}^M$ typically follow a Bernoulli distribution:
\begin{align}\label{eq:1}
f_{ql}(Y_{ij}^m):= P(Y_{ij}^m|z_i = q, z_j = l) \sim Bern(\mu_{ql}),\; \text{for} \;i,j\in \{1,\cdots,N\},\; q,l = 1,\cdots,K,
\end{align}
where $\mu_{ql}$ is the probability of nodes $i$ and $j$ being connected. 

For the heterogeneous stochastic blocks model, the marginal mean $\mu_{ql}$ for each block in the $m$th network can be modeled as a logistic model to incorporate heterogeneity among edges: 
\begin{align}\label{eq:2}
\mu_{ql}^m = exp(\beta_{ql}x^m_{ij})/\big\{1+exp(\beta_{ql}x^m_{ij})\big\},
\end{align}
where $\{(x_{ij}^m)_{N\times N}\}_{m=1}^M$ are edge-wise covariates, and edges within the same community preserve homogeneity by sharing a block-wise parameter $\beta_{ql}$. The joint likelihood function can be decomposed into a summation of edge-wise terms following the conditional independence assumption: 
\begin{align}\label{eq:3}
log P(\bm{Y;Z}) =\displaystyle\sum_{m=1}^M\sum_{q=1}^K\sum_{i=1}^NZ_{iq}log \alpha_{q} + \displaystyle\sum_{m=1}^M\sum_{q,l=1}^K\sum_{{i<j}}^NZ_{iq}Z_{jl}f_{ql}(Y_{ij}^m; \beta_{ql}).
\end{align}
The latent membership $\bm{Z}$ is estimated by $E(\bm{Z|Y})$ through the maximum likelihood estimator of model parameters $\Theta = \{\beta_{ql}; q,l = 1,\cdots, K; \;\alpha_{q}; q = 1,\cdots, K\}$ in (\ref{eq:3}). However, the classical EM algorithm is not applicable here, because the conditional distribution $ \displaystyle P(\bm{Z|Y}) = \frac{P(\bm{Y;Z})}{\sum_{\bm{Z}}P(\bm{Y;Z})}$ becomes intractable in the expectation step.

The variational EM algorithm   
 \citep{mariadassou2010uncovering, jaakkola200110} is one of the  most popular inference methods, and can be applied to approximate the likelihood $P(\bm{Z|Y})$ by a complete factorized distribution $R(\bm{Z}, \bm{\tau}) = \displaystyle \prod_{i=1}^N h(Z_i; \tau_i)$, where $h(\cdot)$ denotes a multinomial distribution, $\bm{\tau}\! =\!(\!\tau_1,\!\cdots,\!\tau_N\!)$ and $\tau_i\! =\!(\!\tau_{i1}, \!\cdots,\!\tau_{iK}\!)$ is a probability vector such that $\sum_{q=1}^K\tau_{iq}=1$. In the expectation step, the likelihood $log P(\bm{Y\!;\! Z})$ is averaged over $R(\bm{Z})$ such that for any $\bm{\tau}$, $E_{R(\bm{Z}, \bm{\tau})}\big\{log P(\bm{Y\!;\! Z})\big\} \leq E_{P(\bm{Z|Y})}\big\{log P(\bm{Y\!;\! Z})\big\}$ where,
 \begin{align*}
E_{R(\bm{Z}, \bm{\tau})}\big\{log P(\bm{Y; Z})\big\}=-&\displaystyle\sum_{m=1}^M\sum_{q=1}^K\sum_{i=1}^N\tau_{iq}log \tau_{iq} + \displaystyle\sum_{m=1}^M\sum_{q=1}^K\sum_{i=1}^N\tau_{iq}log \alpha_{q} + \\& \displaystyle\sum_{m=1}^M\sum_{q,l=1}^K\sum_{{i<j}}^N\tau_{iq}\tau_{jl}f_{ql}(Y_{ij}^m).
  \end{align*} 
Instead of directly maximizing $E_{P(\bm{Z|Y})}\big\{log P(\bm{Y; Z})\big\}$, the variational EM approach alternatively maximizes its lower bound $E_{R(\bm{Z}, \bm{\tau})}\big\{log P(\bm{Y; Z})\big\}$ over model parameters $\Theta$ and variational parameters $\bm{\tau}$, and clusters nodes by $\bm{\tau}$ through
$\hat{z}_i = \argmax_k\{\hat{\tau}_{ik}, k =1,\cdots,K\}$.

Throughout this paper, we consider the conditional version of SBM (CSBM) \citep{bickel2009nonparametric, rohe2011spectral, choi2012stochastic}, where the true membership $\bm{Z^*}$ is fixed. The conditional stochastic block model framework assumes conditional independence among edges, i.e., $Y^m_{i_1j_1}$ and  $Y^m_{i_2j_2}$ are independent given nodes' membership  $z_{i_1}, z_{i_2}, z_{j_1}, z_{j_2}$, and the corresponding log-likelihood of observed sample networks is:
\begin{align}\label{eq:4}
log L_{ind}(\bm{Y|Z}) = \displaystyle\frac{1}{M}\sum_{m=1}^M\sum_{q,l=1}^K\sum_{{i<j}}^NZ_{iq}Z_{jl}\Big\{y^m_{ij}log\;\mu_{ql}+(1-y^m_{ij})log\;(1-\mu_{ql})\Big\}.
\end{align}
The above log-likelihood can serve as a discriminant function in clustering membership $\bm{Z}$ in that if $log L_{ind}(\bm{Y|Z_1}) > log L_{ind}(\bm{Y|Z_2})$ given two membership assignments $\bm{Z_1}$ and $\bm{Z_2}$, then $\bm{Z_1}$ is preferred over $\bm{Z_2}$, since the likelihood for the observed sample networks is higher. 
Naturally, $\bm{Z^*}$ can be estimated by $$\hat{\bm{Z}} = \underset{\bm{Z}}{\mathrm{argmax}}\;log P_{ind}(\bm{Y|Z}).$$
The SBM in (\ref{eq:4}) allows one to differentiate within-community and between-community nodes via utilizing only the marginal information, in that the average connectivity rates within-communities are higher than those between-communities. However, the underlying conditional independence assumption among edges is too restrictive and practically infeasible. In most community detection problems it is common that edges within communities are more correlated. For example, social connections among friends are highly correlated in social networks. However, the dependency among edges is not captured by the traditional SBM, which could lead to significant information loss of the community structure.

\section{Methodology}
   
\subsection{Community detection with dependent connectivity} 

In this paper, we incorporate within-community correlation to improve accuracy and efficiency in identifying communities, in addition to utilizing the edges' marginal mean information, since within-community dependency contains additional information regarding the membership of nodes. This is especially effective when the marginal mean is not informative in differentiating between and within communities' connectivity. 

In this section, we propose an approximate likelihood function to capture the dependency among within-community edges. We assume that each observed sample network $Y^m_{n\times n}$ is generated from an underlying joint binary distribution $P(Y^m)$ such that the correlation among within-community edges is nonnegative. Specifically, for the underlying distribution, the correlation among edges $Y^m_{i_1j_1},Y^m_{i_2j_2}$ within a community satisfies:          
 $corr(Y^m_{i_1j_1},Y^m_{i_2j_2}) = \rho_{q}(i_1,i_2,j_1,j_2) \in [0,1]$ given nodes $z_{i_1}$, $z_{i_2}$, $z_{j_1}$ and $z_{j_2}$ are in the same community $q$, where $1\leq i_1<j_1\leq N, 1\leq i_2<j_2\leq N, (i_1, j_1)\neq (i_2, j_2)$ and $q = 1,\cdots,K$. Note that correlations among each pair of edges could be different. The reason we consider nonnegative correlations among within-community edges is that this reflects the concordance among within-community edges in practice. Equivalently, the edges in community $k$ show concordance only when $$\sum_{i<j; u<v}^NZ_{ik}Z_{jk}Z_{uk}Z_{vk}\hat{y}_{ij}^m\hat{y}_{uv}^m \geq 0,$$
where $\hat{y}_{ij}^m$ and $\hat{y}_{uv}^m$ are normalized binary variables through the marginal mean. Note that positive correlation among edges has been considered for community detection on multiple networks. For example, \citep{paul2018random,pavlovic2015generalised} utilize random effects to model the heterogeneity of the connectivity rate for an individual network, which infers a positive correlation among the edges within the same community. In practice, it has been demonstrated that assuming nonnegative correlations among edges is more sensible and interpretable. For example, the positive pairwise correlation among edges is more likely to produce star or triad relations which are widely observed in social networks \citep{robins2007recent, robins2007introduction}.

\subsection{Approximate Likelihood}

In this section, we propose an informative approximation of the true log-likelihood to cluster $\bm{Z}$ via incorporating interactions among edges within a community in addition to marginal mean information. This is because the exact joint likelihood function of correlated binary distribution $P(Y^m)$ is computationally intractable. Specifically, we construct an approximate likeihood as a substitute of the true likelihood by facilitating the Bahadur representation \cite{bahadur1959representation}. That is, we retain the low-order dependency information among edges within-communities and discard the high-order dependency for computational efficiency. Although the approximate likelihood is not a true likelihood, it still serves the purpose of estimating the membership of nodes.

Consider $T$ dependent binary random variables, then the joint likelihood can be represented through the Bahadur representation: 
\begin{align}\label{eq:5}
P(Y_1=y_1,\cdots,Y_T=y_T) &=\prod_{j=1}^T\mu_{j}^{y_j}(1-\mu_{j})^{1-y_{j}}\big[1 +\sum_{1\leq j_1<j_2\leq T}\rho_{j_1j_2}\hat{y}_{j_1}\hat{y}_{j_2}+ \nonumber \\
&\sum_{1\leq j_1<j_2<j_3\leq T}\rho_{j_1j_2j_3}\hat{y}_{j_1}\hat{y}_{j_2}\hat{y}_{j_3}+\cdots+\rho_{12\cdots T}\hat{y}_{1}\hat{y}_{2}\cdot\cdot\hat{y}_{T}\big],
\end{align}
where 
\begin{align}\label{eq:6}
\mu_{j} = E(Y_j),\; \hat{y}_j = \frac{y_j - E(y_j)}{\sqrt{E(y_j)(1-E(y_j))}},
\end{align}
and
$$\rho_{j_1j_2} = E(\hat{y}_{j_1}\hat{y}_{j_2}),\;\rho_{j_1j_2j_3} = E(\hat{y}_{j_1}\hat{y}_{j_2}\hat{y}_{j_3}),\cdots, 
\rho_{12\cdots T} = E(\hat{y}_{1}\hat{y}_{2}\cdot\cdot\hat{y}_{T}).$$
The idea of Bahadur representation is to approximate the joint distribution of dependent binary random variables as a function of moments with a sequential order. For the community detection problem, the binary random variables represent within-community edges, and the corresponding joint distribution can be explicitly decomposed into a marginal part and a correlation part. The marginal part consists of all the marginal mean $\mu_{ij}$ for each edge, which can be directly modeled through the dependency of the mean on covariates as in (\ref{eq:2}). The correlation part consists of interactions among all possible pairwise-associations of normalized edges, which add correlation information beyond a conditional independence likelihood model. Note that the conditional independence model is a special case of the proposed model when the correlation is zero, and the corresponding Bahadur representation collapses to a marginal part only, which is equivalent to the $\mbox{log} L_{ind}(\bm{Y|Z})$ in (\ref{eq:4}).

There are two major challenges in applying the Bahadur representation to model the interactions among within-community edges. First, the dimension of correlation parameters could be high if all the high-order interactions in (\ref{eq:5}) are incorporated, and this could lead to an increasing computational demand as the size of community grows. To solve this problem, we retain all the second-order interactions, but ignore interactions for higher orders beyond the second order, since the pairwise interactions among edges could be most important. In addition, we can further reduce the number of parameters via a homogeneous correlation structure such that all the pairwise correlations in each community are assumed to be the average within-community correlation, which can be simplified as an exchangeable correlation structure. The rationales of this simplification are based on the following. First, the pairwise correlation parameter $\rho_q(i_1, i_2, j_1, j_2)$ is a nuisance correlation parameter to enhance clustering. Second, both the numerical experiments and theoretical findings show that the density of pairwise correlation among within-community edges plays a more important role than the intensity of the correlation in affecting clustering performance.

The second challenge is that the range of the correlation coefficient could be constrained by the marginal means \citep{diggle1994informative}. Consequently, the correlation parameter space is more restrictive if the variability of marginal means among edges is large. Nevertheless, our primary goal is to construct an objective function which can incorporate information from the marginal mean and correlations of edges within-community, and the objective function is not necessarily the true likelihood function. In the proposed method, we instead construct an approximate likelihood which is more flexible for incorporating highly dependent communities while still achieving computational efficiency.

Specifically, we construct an approximate likelihood $\tilde{L}(\bm{Y|Z})$ incorporating correlated within-community edges as follows:
\begin{align}\label{eq:7}
\mbox{log} \tilde{L}(\bm{Y|Z}) = &\displaystyle\frac{1}{M}\Bigg\{\sum_{m=1}^M\sum_{q,l=1}^K\sum_{{i<j}}^NZ_{iq}Z_{jl}\Big\{y^m_{ij}\mbox{log}\;\mu_{ql}+(1-y^m_{ij})\mbox{log}\;(1-\mu_{ql})\Big\}   \nonumber \\&+ \sum_{m=1}^M \mbox{log} \Big\{1+\sum_{k=1}^K\frac{\rho_{k}}{2}\!\max\{\!\sum_{\substack{i<j;u<v\\(i,j)\neq(u,v)}}^NZ_{ik}Z_{jk}Z_{uk}Z_{vk}\hat{y}_{ij}^m\hat{y}_{uv}^m, 0\} \Big\}\Bigg\},
\end{align}
where $\mu_{ql}$ and $\hat{y}_{ij}^m$ are formulated in (\ref{eq:2}) and (\ref{eq:7}), and $\rho_k$ is the average of pairwise correlation in community $k$.
Notice that the first term in (\ref{eq:8}) is the same as the marginal mean model, and the second term in (\ref{eq:8}) measures the concordance among edges within communities clustering $\bm{Z}$. 

We denote the second term of (\ref{eq:8}) as
\begin{align}\label{eq:8}
\mbox{log} L_{cor}(\bm{Y|Z}) = \displaystyle\frac{1}{M}\Bigg\{\sum_{m=1}^M \mbox{log} \Big\{1+\sum_{k=1}^K\frac{\rho_{k}}{2}\max\{\!\sum_{\substack{i<j;u<v\\(i,j)\neq(u,v)}}^NZ_{ik}Z_{jk}Z_{uk}Z_{vk}\hat{y}_{ij}^m\hat{y}_{uv}^m, 0\} \Big\}\Bigg\}.
\end{align}
Compared with $\mbox{log} L_{ind}(\bm{Y|Z})$ in (\ref{eq:4}), the proposed $\mbox{log} \tilde{L}(\bm{Y|Z})$ has more discriminative power over $\bm{Z}$, since it utilizes more information of the observed dependency within communities corresponding to clustering $\bm{Z}$. In addition, the nonnegativity of $\mbox{log} L_{cor}(\bm{Y|Z})$ ensure the fact that $\mbox{log} \tilde{L}(\mathbf{Y|Z}) \geq \mbox{log} L_{ind}(\mathbf{Y|Z})$ is guaranteed, which implies that adding additional correlation information among edges can be more informative given within-community correlation exists. This leads to higher classification accuracy and estimation efficiency through maximizing (\ref{eq:8}). 

The key part of the proposed method is to predict memberships of nodes through the Bayes factor constructed by the proposed $\mbox{log} \tilde{L}(\bm{Y|Z})$. Suppose the memberships of other nodes $\bm{Z_{-i}}$ are known, then we classify node $i$ based on the following Bayes factor:
$$\frac{\tilde{L}(\mathbf{Y}|\bm{Z_{-i}}, Z_{iq}=1)}{\tilde{L}(\mathbf{Y}|\bm{Z_{-i}}, Z_{ik}=1)} = exp\Big\{\mbox{log} \tilde{L}(\mathbf{Y|\bm{Z_{-i}}}, Z_{iq}=1) - \mbox{log} \tilde{L}(\mathbf{Y|\bm{Z_{-i}}}, Z_{ik}=1)    \Big\}.$$
If the above Bayes factor $>1$, then the probability of node $i$ in community $q$ is larger than that of community $k$. The Bayes factor can be further decomposed as:
\begin{align}\label{eq:9}
\frac{\tilde{L}(\mathbf{Y}|\bm{Z_{-i}},Z_{iq}=1)}{\tilde{L}(\mathbf{Y}|\bm{Z_{-i}},Z_{ik}=1)} = \frac{L_{ind}(\mathbf{Y}|\bm{Z_{-i}},Z_{iq}=1)}{L_{ind}(\mathbf{Y}|\bm{Z_{-i}},Z_{ik}=1)}\frac{L_{cor}(\mathbf{Y}|\bm{Z_{-i}},Z_{iq}=1)}{L_{cor}(\mathbf{Y}|\bm{Z_{-i}},Z_{ik}=1)},
\end{align}
which contains both the marginal ratio and the correlation ratio. It is clear that when the marginal information is weak in differentiating two communities, the marginal ratio is close to 1, and if the correlation ratio is informative, it can enhance the Bayes factor to improve community detection. In addition, the correlation ratio also serves as a correction to lower the estimation bias. 

\begin{figure}[H]
\centering
   \includegraphics[width=5.5in, height=2.6in]{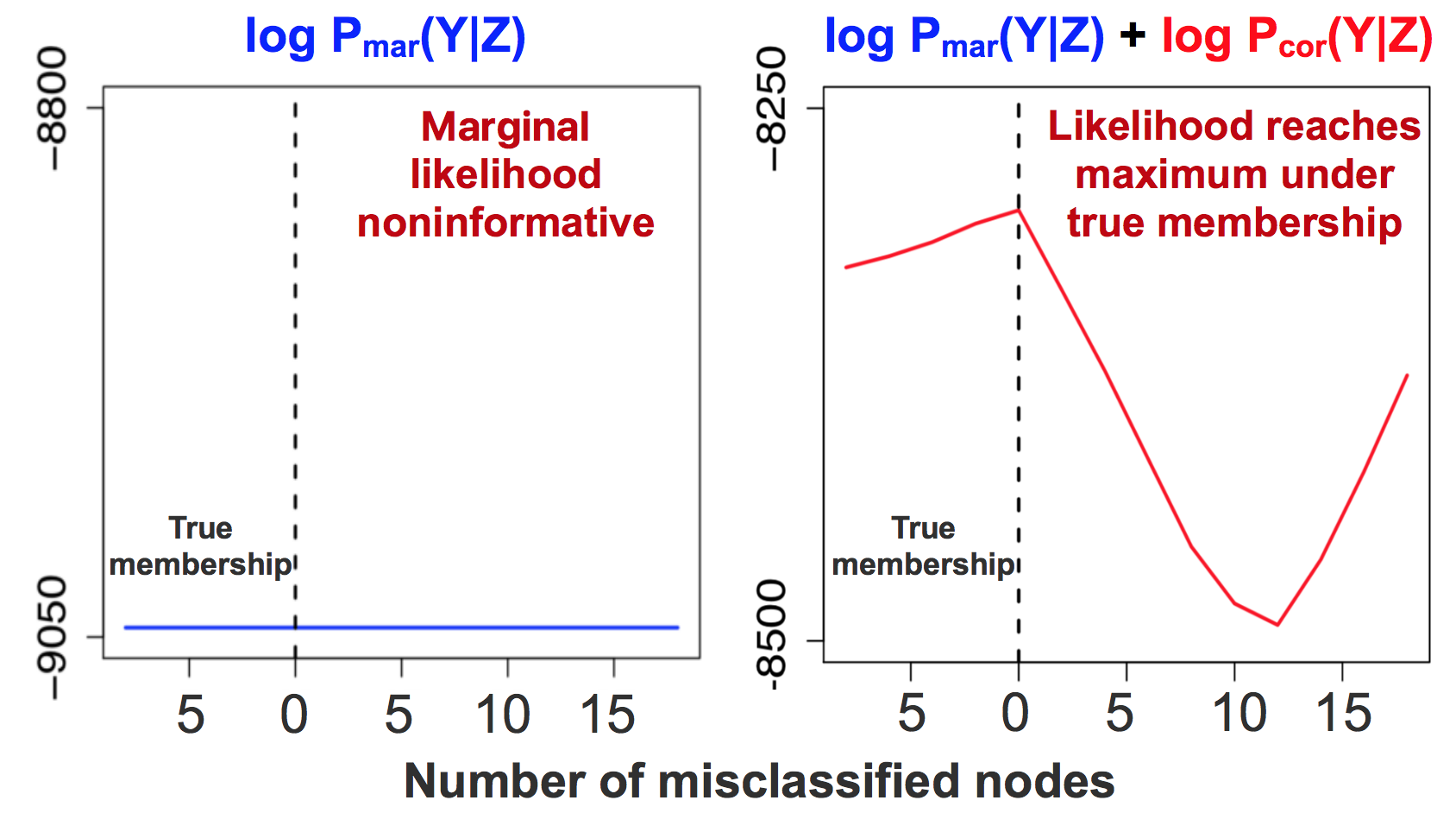}
   \vskip -0.5ex
  \caption{Likelihood of multiple networks with 30 nodes from two communities. \textit{Left}: Traditional SBM likelihood. \textit{Right}: The proposed pseudolikelihood incorporating correlation information.}
  \label{fig:1}
\end{figure}

We illustrate the advantage of the proposed method in (\ref{eq:8}) over the conditional independent likelihood (\ref{eq:4}) using a simple numerical illustration. Specifically, we generate multiple networks based on the SBM with 30 nodes evenly split between two communities. The marginal means of within-community and between-community edges are the same at 0.5, implying that the marginal mean is not informative. We assume a true exchangeable correlation $\rho = 0.6$ for within-community edges.
Figure \ref{fig:1} 
illustrates that the likelihood function changes as memberships of nodes change with some misclassified nodes. The left graph is based on the conditional independent SBM utilizing only marginal information, which does not differentiate the two communities at all due to weak marginal information. However, the proposed approximate likelihood in the right graph has high differentiation power for the nodes' memberships, and reaches maximum when the true memberships are selected. 

\section{Algorithm and Implementation}
In this section, we propose a two-step algorithm to maximize the proposed approximate likelihood function. In addition, we provide implementation strategies to improve the stability and efficiency of the algorithm.
\subsection{Algorithm} 
To estimate the true membership $\bm{Z^*}$ of nodes, we can ideally search through all the possible $\bm{Z}$ and choose the one with the largest 
$\mbox{log} \tilde{L}(\bm{Y|Z})$. However, this becomes infeasible when the number of nodes $N$ and the number of communities $K$ increases.  In the following, we propose an iterative two-step algorithm to maximize $\mbox{log} \tilde{L}(\bm{Y|Z})$ in (\ref{eq:7}). 
\begin{table}[H]
\begin{tabular*}{1\linewidth}{l}
\hline\hline
\textbf{Algorithm 1}\\
\hline
\addlinespace[1ex]
\textbf{Step 1}: Input an initial membership probability for each node: $\alpha_{iq}^{(0)}, \; 1\leq i \leq N,\;1\leq q\leq K$\\ \hspace{14mm}through spectral clustering on individual sample networks \\
\addlinespace[1ex]
\textbf{Step 2}: At the $s$th iteration, given $\{\beta_{ql}^{(s-1)},\,\rho_{q}^{(s-1)}\}_{q,l=1}^K$ and $ \{\alpha_i^{(s-1)}\}_{i=1}^N $ from the $(s-1)$th iteration:\\[5pt] 
 \hspace{0.14in}  \textbf{(i) Maximization:} block-wise update $\beta_{ql}^{(s)}$ and $\rho_q^{(s-1)},\; q,l =1,\cdots,K$; \\[5pt]
\hspace{0.27in} (a) Obtain $\beta_{ql}^{(s)}$ through GEE with current membership as working correlation \\  
      \addlinespace[0.7ex]
     
     \hspace{0.31in}(b) Update $\rho_{q}^{(s)}$ through the method of moments estimator given current membership  \\
      

    
\addlinespace[2ex]
\hspace{0.12in}   \textbf{(ii) Expectation:} given  $\{\beta_{ql}^{(s)},\;\rho_{q}^{(s)}\}_{q,l=1}^K$ , update $\{\alpha_i^{(s)}\}_{i=1}^{N}$: \\
   \addlinespace[1ex] 
  \multicolumn{1}{c}{$\alpha_{iq}^{(s)} = \frac{\alpha_{iq}^{(s-1)}\tilde{L}(\mathbf{Y}|\alpha_{-i}^{(s-1)},Z_{iq}=1)}{\sum_{k=1}^K\alpha_{ik}^{(s-1)}\tilde{L}(\mathbf{Y}|\alpha_{-i}^{(s-1)},Z_{ik}=1)},\;i = 1,\cdots,N, \;q = 1,\cdots,K.$}\\[8pt]
\textbf{Step 3}: Iterate until $\max\limits_{1\leq i\leq N} |\alpha_{i}^{(s)}-\alpha_{i}^{(s-1)}|<\epsilon$. \\
\addlinespace[1ex]
\textbf{Step 4}: Obtain the membership $z_i$ of clusters by\\ 
\addlinespace[1ex] 
\hspace{1in}$\{\alpha_{i}^{(s)}\}_{i=1}^N$: $z_i = \max_{k}\{\alpha^{(s)}_{i1},\cdots,\alpha^{(s)}_{iK}\},\;i =1,\cdots,N$.\\
\addlinespace[0.5ex]
\hline\hline
\end{tabular*}
\end{table}

Here we directly maximize the approximate likelihood instead of a true likelihood as in the EM algorithm.
In the expectation step, we alternatively update membership of each node while fixing other nodes, where $\tilde{L}(\mathbf{Y}|\alpha_{-i};Z_{ik})$ has the same formulation as $\tilde{L}(\bm{Y|Z})$ in (\ref{eq:7}) with $\{Z_{iq}\}_{N\times K}$ replaced by its expectation $\{\alpha_{iq}\}_{N\times K}$, except $Z_{ik}$. Note that $\alpha_{iq}$ is not the expectation under the true underlying joint  distribution $P(Y, Z) = P(Y|Z)P(Z)$. Instead, it corresponds to the distribution defined by the approximate likelihood in (\ref{eq:8}).
In the expectation step, the memberships are updated through the Bayes factor in (\ref{eq:9}) with the proposed $\tilde{L}(\mathbf{Y|Z})$. In the maximization step, we estimate the community-wise parameters $\beta_{ql}$ through the generalized estimating equation where the working correlation is exchangeable structure given the current membership of nodes and estimated average correlation $\rho_q$.  
Note that the variational EM is a special case of the proposed algorithm if the correlation information is ignored and the conditional independent model in (\ref{eq:4}) is assumed.

%

\subsection{Computation and Implementation:} 

To ensure computational stability, the community-wise parameters $\beta_{ql}$ could be estimated through a simplified generalized estimation equation assuming an independent working correlation in algorithm 1.  This is because the primary interest of community detection is classification accuracy, and the empirical studies show that correlation information plays a relatively minor role in parameter estimation. 

We can achieve a better approximation to the true likelihood if higher-order moments are incorporated in the Bahadur representation in (\ref{eq:6}), which also increases its discrimination power. However, higher-order correlation could also increase the computational cost. Alternatively, we can recover partial higher-order interactions (e.g., the fourth order) derived from low order interactions (e.g., the second order). For example, consider four normalized edges 
$\hat{Y}^m_{i_1j_1}, \hat{Y}^m_{i_2j_2}, \hat{Y}^m_{i_3j_3}$ and $\hat{Y}^m_{i_4j_4}$ within the same community $k$ with a  positive fourth order correlation among them, we have
\begin{align}\label{eq:10}
E\big(\hat{Y}^m_{i_1j_1}\hat{Y}^m_{i_2j_2}\hat{Y}^m_{i_3j_3}\hat{Y}^m_{i_4j_4}\big)\geq E\big(\hat{Y}^m_{i_1j_1}\hat{Y}^m_{i_2j_2})E\big(\hat{Y}^m_{i_3j_3}\hat{Y}^m_{i_4j_4}) = \rho_{i_1j_1i_2j_2}\rho_{i_3j_3i_4j_4}.
\end{align}
To simplify notation, denote $(Z_{1k}Z_{2k}\hat{Y}^m_{12},Z_{1k}Z_{3k}\hat{Y}^m_{13},\cdots,Z_{2k}Z_{3k}\hat{Y}^m_{23},\cdots,Z_{(N-1)k}Z_{Nk}\hat{Y}^m_{(N-1)N})$ as $(\gamma_{1}^m,\gamma_{2}^m,\cdots,\gamma_{N_0}^m)$, where $N_0 = \frac{N^2-N}{2}$. Then the second-order interaction term for the community $k$ in $L_{cor}(\bm{Y|Z})$ is
\begin{align*}
\frac{\rho_{k}}{2}\sum_{\substack{i<j,u<v\\(i,j)\neq (u,v)}}^NZ_{ik}Z_{jk}Z_{uk}Z_{vk}\hat{y}_{ij}^m\hat{y}_{uv}^m  = \rho_k\sum_{s<t}^{N_0}\gamma_{s}^m\gamma_{t}^m.
\end{align*}
Based on (\ref{eq:11}) and given $\bm{Z}$, we can approximate the fourth-order interaction for community $k$ under the exchangeable correlation structure by its lower bound:
\begin{align}\label{eq:11}
\sum_{\substack{s_1<t_1,s_2<t_2\\(s_1,t_1)\neq (s_2,t_2)}}^{N_0}\!\!\!\!\!\!\!\!\!\!\!\frac{E(\gamma_{s_1}^m\gamma_{t_1}^m\gamma_{s_2}^m\gamma_{t_2}^m)}{2}\gamma_{s_1}^m\gamma_{t_1}^m\gamma_{s_2}^m\gamma_{t_2}^m \geq\!\!\!\!\!\!\!\!\!\!\! \sum_{\substack{s_1<s_2,t_1<t_2\\(s_1,t_1)\neq (s_2,t_2)}}^{N_0}\!\!\!\!\!\!\frac{\rho_k^2}{2}\gamma_{s_1}^m\gamma_{t_1}^m\gamma_{s_2}^m\gamma_{t_2}^m\! =\!\!\Big(\rho_k\sum_{s<t}^{N_0}\gamma_{s}^m\gamma_{t}^m\Big)^2\!\!\!\!- \!\!\rho_k^2\sum_{s<t}^{N_0}(\gamma_{s}^m\gamma_{t}^m)^2.
\end{align}
Note that the above lower bound of the fourth-order interaction can be calculated by the 
second-order interaction term in $L_{cor}(\bm{Y|Z})$. Therefore, we can still incorporate higher-order terms in $log \tilde{L}(\bm{Y|Z})$ without additional computational cost. For other types of non-exchangeable correlation structures, we can incorporate partial higher-order correlation similarly as above. The main difference is that each pair of edges is associated with a specific correlation given a dependency structure. Therefore, the simplified lower bound for higher-order correlations such as (\ref{eq:12}) does not hold in general, and could have a more complex form depending on the specific correlation structure.

In the following, we also provide some guidelines for determining the number of communities $K$ and initial membership of nodes. For a single network, the criterion-based methods choose $K$ to maximize a certain probabilistic criterion such as the integrated likelihood \citep{geng2018probabilistic, daudin2008mixture, latouche2012variational}, composite likelihood BIC \citep{saldana2017many} or modularity criterion \citep{blondel2008fast}. In addition, spectral methods estimate $K$ through the spectral property of the transformed adjacent matrix, such as a Laplacian matrix \citep{newman2006finding}, non-backtracking matrix \citep{bordenave2015non} or Bethe Hessian matrix \cite{saade2014spectral}. In the hierarchical Bayesian framework, the number of communities is treated as a model parameter given a certain prior distribution and is jointly estimated with nodes' memberships using the MCMC \citep{geng2018probabilistic, newman2016estimating, nobile2007bayesian}. For multiple networks, we can extend the above techniques to estimate a consensus number of communities combining observed realizations of the SBM from each individual network. 

In the context of the proposed within-community dependent modeling, we can first perform the modularity-maximizing method or spectral clustering on each individual network to obtain $K$, then take the average of these individual estimated $K$, which can be treated as a consensus number of communities. The above procedure is sensible under two considerations. First, each sample network is a realization of the SBM so that the individual estimation of $K$ is randomly distributed around the true underlying $K$. Thus the average of individual estimations provides an estimation of $K$ with low-bias and low-variance. Second, the spectral clustering or modularity methods are more favorable than other methods, due to their relatively low computational cost in estimating $K$. This is especially effective when the sample size of networks is large.

As an EM-type algorithm, the proposed optimization procedure can only guarantee the local maximum and requires multiple initializations to find the global maximum. In this paper, we obtain the membership initializations through spectral clustering on  different sample networks, a benchmark algorithm for the traditional SBM. Spectral clustering is a model-free clustering algorithm and is able to provide a warm start for nodes' memberships.




\section{Theoretical Results}
In this section, we establish the consistency of the estimated nodes' membership based on  the independent likelihood and the approximate likelihood approaches. In addition, we provide the computational convergence theorem for the proposed iterative algorithm in section 4.  
Compared to the independent likelihood approach, we show that the approximate likelihood approach leads to a computationally faster convergence rate regarding nodes' membership estimation.

\subsection{Consistency of nodes' membership estimation}
In this subsection, we study the consistency of the maximization likelihood estimator for both the independent likelihood and the approximate likelihood at the population level. With the independence assumption among within-community edges, the consistency and convergence rate of the MLE estimator can be obtained by \citep{celisse2012consistency, zhang2017theoretical}. However, the convergence property of the MLE remains unknown if there exists a local dependence among edges.

One significant distinction using the independence assumption if the edges are correlated is that the increasing number of nodes and number of edges do not necessarily guarantee a lower misclassification rate and computationally faster convergence. This is because the discrepancy between marginal means from within-community and between-community is not accumulated due to the pairwise correlation, though it can be accumulated through increasing the number of sample networks. However, we show that the proposed approximated approach is able to benefit from the increasing number of nodes, and therefore achieves a faster computational  convergence compared to the independent likelihood approach.

Without loss of generality, we assume that $x_{ij}^m = 1, m = 1,\cdots, M, i,j = 1,\cdots, N$. That is, all the edges within the same block have the same marginal mean such that $\mu_{z_iz_j}: = E(Y_{ij}^{m}|i\in q, j\in l) = \frac{\exp(\beta_{ql})}{1+\exp(\beta_{ql})}$. We denote that the true marginal mean as $\Theta = \{\mu_{ql}, 1\leq q<l \leq K\}$. 
The following two regularity conditions regarding identifiability are standard:

(C1). Suppose for every $q \neq q', 1\leq q,q'\leq K$ and $l\neq l', 1\leq l,l'\leq K$,  we have
$ \mu _ { ql } \neq \mu _ { q ^ { \prime }l' }$. In addition, all the $\mu_{ql}$ are bounded such that $c_2\leq \mu_{ql}\leq c_1, q,l =1,\cdots,K$ where $0<c_2\leq  c_1<1$.

(C2). Community sizes from all sample networks are bounded above and below by $\kappa_1N \leq |\{i\in \{1,2,\cdots,N\}: Z^*_{iq} = 1\}|\leq \kappa_2N,\; q=1,\cdots,K$, where $\kappa_1$ and $\kappa_2$ are constants such that $0<\kappa_1<\kappa_2<1$.

In the following, we establish the consistency of membership estimation for both the independent likelihood approach and the proposed approximate likelihood approach. For the within-community edges,  we define the edgewise second-order pairwise correlation density as
\begin{align}
\label{eq:12}
\lambda = \lambda_{ij}^m := \frac{\#|\{(u,v):cor(Y^m_{ij}, Y^m_{uv})>0, Z_u = Z_v = k\}|}{N_k(N_k-1)/2-1}\; \text{for edge $Y_{ij}^m$ in community k}
\end{align} 
where $k = 1,2,\cdots,K$ and $N_k(N_k-1)/2-1$ is the number of edges within community $k$ for the sample network $\bm{Y}^m$. For simplicity, we assume the homogeneous second-order correlation density such that $\lambda_{ij}^m = \lambda$ for all the within-community edges. Here $\lambda\in [0,1]$ determines the intensity of local dependency within a community. Specifically, $\lambda = 0$ indicates that within-community edges are all independent, while $\lambda = 1$ indicates that all the within-community edges are pair-wisely correlated. In addition, correlation density $\lambda$ is allowed to depend on the number of nodes, and increases such that it can model a more general class of correlation structure. For example, in a hub structure, an edge is only correlated with those sharing the same hub nodes and the density $\lambda = \frac{N_k-1}{N_k^2-1} = O_N(\frac{1}{N_k})$.

To establish asymptotic consistency for the proposed likelihood, we assume the sparsity of high-order correlation among within-community edges.

(C3). The number of third and fourth-order correlations defined in (\ref{eq:5}) among within-community edges do not exceed the order of the size of second-order correlations. Specifically, for edge $Y^m_{ij}$ in community $k$, $\#|\{(i,j), (u_1, v_1), (u_2, v_2) :E(\hat{Y}_{ij}\hat{Y}_{u_1v_1}\hat{Y}_{u_2v_2})>0\}|\leq O_N(\lambda(N_k^2))$. In addition, $\#|\{(i,j), (u_1, v_1), (u_2, v_2), (u_3, v_3): E(\hat{Y}_{ij}\hat{Y}_{u_1v_1}\hat{Y}_{u_2v_2}\hat{Y}_{u_3v_3})>0\}|\leq O_N(\lambda(N_k^2)), k = 1,2,\cdots,K$. 

In general, assume that the pairwise correlations among the within-community edges are sufficient to cover a broad class of Markov dependence modeling under the general exponential random graph model. This includes the most commonly used edge dependence configurations such as a star, a triangular shape subnetwork \citep{newman2003structure} and the k-triangles shape \citep{pattison20029}. Although considering that the additional higher-order edge correlation improves the model's complexity, it could increase higher computational cost and instability. Empirically, it is sensible to assume that higher-order correlation only exists when second-order correlation already exists among edges, for the sake of identifiability and interpretability of the model. Otherwise, it could lead to the 'near degeneracy' \citep{handcock2003statistical} when a higher-order dependency masks a lower-order dependency.

Let $P_{Z^*} := \mathbb{P}(\cdot | Z = z^*; \Theta)$ denote the conditional distribution of edges given the true membership of nodes and true parameters.

\begin{theorem}
\label{T3.1}
Under the regularity conditions (C1)-(C3), we establish the convergence rate of the membership estimator $z$ using the independent likelihood approach. That is, for every $t>0$ and $z \neq z^*$,
\begin{align}\label{t3.1}
P _ { Z* }\Big\{ \frac { L_{ind}(\bm{Y}| \bm{Z}= z; \Theta)} { L_{ind}(\bm{Y}| \bm{Z }= z^*; \Theta)} > t \Big\} =\mathcal{O}(\exp\Big\{-C_1\frac{rNM}{1+\rho\kappa_2 N\min(r,\kappa_2\lambda N)}\Big\}),
\end{align}
where $r = \|z-z^*\|_0$ is the number of misclassified number of nodes up to the permutation of the label, and $\rho$ is the largest pairwise correlation among within-community edges. In addition, $C_1 = \frac{c}{\max\{\log \frac{c_1}{c_2}, \log\frac{1-c_2}{1-c_1}\}}$, where $c$ is a positive constant.  
\end{theorem}

For the independent likelihood approach, the convergence rate depends on the number of sample network $M$ and the density of the pairwise correlation among within-community edges $\lambda$. The probability of true membership goes to 1 as $M$ or the node size $N$ increases given a relatively sparse pairwise correlation such that $\lambda N = o_N(1)$. If there is no pairwise correlation among edges, hence $\lambda=0$, then the convergence rate increases to $O_{N,M}(\exp(-C_1rNM))$, which degenerates in the conditional independent setting to the convergence rate established in \citep{celisse2012consistency}.

In the case of the exchangeable correlation structure for within-community edges, hence $\lambda=1$, the convergence rate decreases to the order of $O_{N,M}(\exp(-C_1\frac{M}{\kappa_2\rho}))$, and therefore does not benefit from the increasing number of nodes. In this case, the consistency relies on accumulating independent sample networks. Theorem \ref{T3.1} also implies that the independent likelihood approach is unable to fully accumulate discriminative power from the increasing number of nodes when there exists dependency among within-community edges. Indeed, the convergence rate of the independent likelihood approach decreases in terms of network size N as the within-community correlation density $\lambda$ increases. 
However, we show that the proposed approximate likelihood approach still benefits from increasing nodes size even under the exchangeable correlation structure among edges.

\begin{theorem}
\label{T3.2}
Under the regularity conditions (C1)-(C3), we establish the convergence rate of the estimator $z$ using the proposed approximate likelihood approach. That is, for every $t>0$, $z \neq z^*$, and \textcolor{blue}{$\lambda >0$},
\begin{align}\label{t3.2}
P _ { Z* }\Big\{ \frac {\tilde{L}(\bm{Y}| \bm{Z}= z; \Theta)} { \tilde{L}(\bm{Y}| \bm{Z }= z^*; \Theta)} > t \Big\} = \mathcal{O}(\exp\Big\{-C\frac{r\lambda NM(1+\lambda N^2)}{1+c\rho \kappa_2 N\min(r, \kappa_2\lambda N)}\Big\}),
\end{align}
where $M>O_N(\frac{1}{\lambda})$, $r = \|z-z^*\|_0$ is the number of misclassified nodes up to the permutation of the label, $C,c$ are positive constants, and $\rho$ is the largest within-community correlation.
\end{theorem}
Given the same number of network $M$ and node size $N$, the proposed approximate likelihood approach is able to achieve a constantly faster convergence rate compared with (\ref{t3.1}) since it has an additional term $\lambda^2 N^3M$ on the numerator in (\ref{t3.2}) given $\frac{1}{\lambda} = o_N(N)$.  
Specifically, the proposed approach is most superior under the exchangeable correlation structure ($\lambda = 1$), where the convergence rate of the independent likelihood is at the order of $O_{N,M}(\exp(-cM))$, in contrast to the proposed convergence rate of $O_{N,M}(\exp(-cNM))$.
Intuitively, incorporating the correlation information increases the effective sample size of within-community edges. Under the sparsity assumption of higher-order correlation among edges, the proposed approach benefits from accumulating information on the second-order interactions among edges, while the independent likelihood approach only accumulates information from the first-order marginal mean of edges.

\subsection{Computational convergence for the proposed algorithm}

In this subsection, we provide the computational convergence property of the proposed algorithm in Section 4. The main difference between the proposed method and the variational EM lies in the Bayes factor of (\ref{eq:9}) in the expectation step from Algorithm 1. If we replace $\tilde{L}(\bm{Y|Z})$ by the conditional independent likelihood $L_{ind}(\bm{Y|Z})$ in (\ref{eq:4}) in the expectation step, the standard variational EM becomes a special case of Algorithm 1.
Notice that \citep{zhang2017theoretical} establishes computational convergence with the minimax rate of misclassification only when the within-community edges are independent. In addition, it assumes that the within-community marginal means are all the same, which is too restrictive in practice.

In the following, we establish the computational convergence for the proposed approximate likelihood. Specifically, we are able to show a faster convergence speed and a lower estimation bias compared to the existing one based on the independent likelihood in \citep{zhang2017theoretical}. The following Theorem \ref{T4.2} also relaxes the homogeneous marginal mean assumption and allows the marginal means from within-community and between-community to be different. We denote the estimated memberships of nodes at the $s$th iteration as $\bm{\alpha^{(s)}} = (\alpha_1^{(s)},\cdots,\alpha_N^{(s)})$ from Algorithm 1. In addition to the assumptions (C1-C3) in Section 5.1, we require two regularity conditions for the following theorems:

(C4). Suppose the distance between initial membership $\bm{\alpha}^{(0)}$ and true membership $z^*$ is bounded: $\|\bm{\alpha}^{(0)} - z^*\|_1 \leq cN^{1-\eta}$ where $0<\eta<1$. 

A common issue for most EM-type algorithms including the one proposed is that they only guarantee convergence to a local optimum. If the likelihood function is unimodal, then the EM-type algorithm converges to the MLE as the unique global optimum. However, the proposed approximate likelihood is non-convex and multi-modal. Therefore, we assume that the initials are in the neighborhood of the MLE to ensure the convergence of the EM algorithm\citep{balakrishnan2017statistical,wu1983convergence}. Condition C4 is a common assumption to guarantee computational convergence for EM-type algorithms \citep{zhang2017theoretical, jain2013low, keshavan2010matrix}. 

(C5). The estimated marginal mean $\hat{\mu}_{ql}$ has a bounded bias from the truth, i.e., 
$0< \gamma_1 \leq \frac{\hat{\mu}_{ql}}{\mu_{ql}} \leq \gamma_2,\;q,l=1\cdots,K, $. 
\begin{theorem}
\label{T4.2}
Under the regularity conditions (C1)-(C5) and given $N$ is sufficiently large, we establish the convergence property of Algorithm 1 via incorporating correlation information. That is, with the correlation density $\frac{1}{\lambda} = o_N(N^{\frac{\eta}{2}})$, as $M$ and $N$ increases with $O_N(\frac{1}{\lambda}) < M\leq o_N(N^{2-\frac{\eta}{2}})$, then
\begin{align}\label{eq:21}
E\|\bm{\alpha^{(s+1)}} - z^* \|_1 \leq c_1NK\exp(-c_2(1+\lambda)MN)+ \frac{c_3N^{1+\frac{\eta}{4}}\|\bm{\alpha}^{s} - \bm{z^*}\|_1}{(1+\lambda  N^{2+\frac{\eta}{4}})M},
\end{align}
where $c_1, c_2, c_3$ are positive constants.
\end{theorem}

In Theorem \ref{T4.2}, the first term on the right side of the inequality represents the irreducible estimation bias which measures the discrepancy between the community structure and its realization. The second term provides a decreasing rate of misclassification along each iteration. Theorem $\ref{T4.2}$ indicates that the estimated memberships are closer to the true memberships compared to the previous iteration step at a rate of $\frac{N^{1+\frac{\eta}{4}}}{(1+\lambda  N^{2+\frac{\eta}{4}})M}$, where a larger sample size $M$ or node size $N$ contribute a faster convergence and a lower estimation bias. In general, Theorem $\ref{T4.2}$ guarantees the convergence of the iterative algorithm even without incorporating correlation information, but improves the convergence rate and estimation bias when correlation information is incorporated.

Specifically, in contrast to the computational convergence rates in Theorem $3.1$ of \citep{zhang2017theoretical}, our Theorem $\ref{T4.2}$ shows that incorporating the correlation information enables us to reduce the estimation bias and accelerate the convergence rate. Specifically, if we consider the $M$ sample networks with node size $N$ as a single network with $MN$ nodes, then the proposed approximate likelihood approach reduces the order of the estimation bias from $O_{N,M}(MN\exp(-cMN))$ in \citep{zhang2017theoretical} to $O_{N,M}(N\exp\{-c'(1+\lambda)MN\})$ in (\ref{eq:21}), and the order of the convergence rate from $O_{N,M}(\frac{1}{\sqrt{MN}})$ in \citep{zhang2017theoretical} to $O_{N,M}(\frac{N^{1+\frac{\eta}{4}}}{(1+\lambda  N^{2+\frac{\eta}{4}})M})$ in (\ref{eq:21}), respectively. Compared with the convergence rate of the membership estimator assuming conditional independence in (\ref{eq:21}) when the correlation density $\lambda = 0$, incorporating within-community  correlation accelerates the computational convergence when there is a sufficiently large number of within-community correlated edges corresponding to $\lambda > \frac{1}{\sqrt{MN}}$. 
\section{Numerical Studies}
In this section, we conduct simulation studies to illustrate the performance of the proposed method on community detection in networks for dependent edges within-community. In particular, we compare our method to the existing variational EM method which assumes conditional independence among edges.

\subsection{Study 1: Networks with dependent within-community connectivity}
In the first simulation study, we consider networks where edges within the same community are correlated and compare the performance of various methods under different network sample sizes with various magnitudes of marginal means for within-community and between-community. 

Suppose the memberships of nodes $\bm{Z^*} = \{\bm{Z}_1,\cdots,\bm{Z}_n\}$ in the networks are given with $K$ communities, where $\bm{Z}_i$ is a binary indicator vector corresponding to the membership of nodes $i$. Conditional on $\bm{Z^*}$, edges in each sample network are generated following the Bernoulli marginal distribution as in (\ref{eq:1}), where within-community edges follow an exchangeable correlation structure as in (\ref{eq:5}).
Here we assume that between-community edges are independent from each other. The block-wise marginal means $\mu_{ql}\;(q,l = 1,\cdots,K)$ are associated with edgewise covariates through (\ref{eq:2}). In addition, the edgewise covariates follow a uniform distribution, where within-communities covariates
\begin{align}\label{eq:14}
x^m_{ij}\sim Unif(a_1,a_2)\;\text{if}\; Z_{iq}=Z_{jq} = 1,
\end{align}
and between-community covariates
\begin{align}\label{eq:15} 
x^m_{ij}\sim Unif(b_1, b_2)\;\text{if}\; Z_{iq} \neq Z_{jq}, q = 1,\cdots,K.
\end{align}
Although the probability of each edge is different, the edges within the same community share the same coefficient $\beta_{ql}$ in (\ref{eq:2}). 
In the following simulation studies, we generate correlated unweighted edges through the R package "MultiOrd."

Specifically, the sample networks consist of 40 nodes split into two communities. In a balanced community network, each community has 20 nodes. In an unbalanced case, two communities are comprised of 10 and 30 nodes, respectively. We compare the performance under different sample sizes of networks with $M=20,\,40$ and 60, and different intensities of within-community dependency with correlation coefficient $\rho = 0$, $0.3$ and $0.6$.

To simulate a weak marginal signal case, we let the block-wise parameters be $\beta_{11}=1,\, \beta_{22} = 1.5$ and $\beta_{12}=\beta_{21}=0$. The means of within-community and between-community covariates are 0 with $a_1=b_1=-0.2$ and $a_2=b_2=0.2$ in (\ref{eq:14}) and (\ref{eq:15}). Here, although the marginal mean of within-community edges is slightly larger than that of between-community edges on average due to the convexity of the logistic link function in (\ref{eq:2}), the marginal means of within-community edges and between-community edges are very close. 

For a strong marginal signal case, the block-wise parameters are $\beta_{11}=0.3,\, \beta_{22} = 0.6$ and $\beta_{12}=\beta_{21}=0.2$. The within-community covariates are generated via (\ref{eq:14}) with $a_1=0.9$ and $a_2=1.1$, and between-community covariates are generated from (\ref{eq:15}) with $b_1= -0.8$ and $b_2 =-0.6$. Note that there is a distinct gap between within-community and between-community marginal means, thus the marginal signal is more dominant for nodes within communities.

We use the Adjusted Rand Index (ARI) to measure the performance of clustering. The ARI takes a value between $-1$ and $1$, where $1$ represents a perfect matching of true memberships and predicted memberships of clustering, $0$ indicates a random clustering and a negative value indicates that the agreement is less than the expectation from a random result. In the following simulations, we choose five fixed initial memberships of nodes in both balanced and unbalanced communities. These initials can be obtained from spectral clustering on sample networks. The Adjusted Rand Indices based on these chosen initials range between $0.30$ to $0.34$ under the unbalanced community case and between $0.25$ to $0.29$ under the balanced community case, which are far from the true memberships.

We compare the performance of clustering and parameter estimation for the proposed method applying the second-order (Bahadur\textsubscript{2nd}) and the fourth-order (Bahadur\textsubscript{4th}) Bahadur approximation, and the variational EM (VEM) approach with only marginal information.

In Table \ref{tab:1} and Table \ref{tab:2}, the proposed method with the second-order and fourth-order approximations outperform the variational EM in clustering. Specifically, under the weak marginal signal case in Table 1, 
the Adjusted Rand Index of the variational EM are 0.34 under different network sizes and correlation strengths, which are similar to the ones calculated by fixed initials. In addition, since the distributions of marginal means from within-community and between-community are similar, the variational EM marginal approach barely improves over the initial memberships as it only utilizes the marginal information. However, the proposed method with the second-order or fourth-order Bahadur representation improves on the ARI by about $280 \%$, compared to the VEM when $\rho=0.3$ and $\rho=0.6$. In addition, the performance of the proposed method improves by $1\sim5\%$ as the number of sample networks increases from 20 to 60. Furthermore, incorporating the fourth-order interaction can slightly improve the accuracy of clustering. 

We notice that when the correlation is as moderate as 0.3, the proposed method still achieves significant improvement over the variational EM and almost fully recovers the true memberships of clustering. We consider this as an intrinsic advantage of the proposed method in capturing the relatively weak dependency among edges to improve the clustering. This is because the proposed method not only captures pairwise dependency but also reflects connectivities among nodes within a community. That is, even a weak dependency among pairwise connectivities can lead to an accumulative information recovery of clustering.

Table \ref{tab:2} illustrates the clustering performance when the marginal signal is strong. In contrast to Table \ref{tab:1}, the variational EM significantly improves on clustering because of the large discrepancy between the within-community marginal mean and the between-community marginal mean. Nevertheless, incorporating the correlation among within-community edges still improves the clustering accuracy by $20 \%$ to $26 \%$ under various sample sizes of networks and intensities of correlation. The clustering accuracy of the proposed method improves when either the sample size or the correlation increases. In general, stronger correlation and a larger sample size lead to better performance when the marginal signal itself is strong.

In addition to clustering, we also provide estimation of the marginal parameters. Tables \ref{tab:6}, \ref{tab:7} and \ref{tab:8} compare parameter estimation between the proposed method and the variational EM when the marginal signal is weak. For within-community parameters $\beta_{11}$ and $\beta_{22}$, the estimation of the proposed method consistently reduces bias $30\sim 99\%$ more than the variational method, except when $M=20$ and $\rho = 0.6$. This is because the sample size $M=20$ is not sufficiently large to offset the high variance among highly-correlated within-community edges. For the between-community parameter $\beta_{12}$, the estimation bias of the proposed method consistently decreases more than $80\%$ compared to the VEM under all settings. Additionally, the standard errors of the proposed estimator decrease faster than the variational method as the sizes of networks increase.

We also investigate the clustering performance of the independent likelihood and the proposed approximate likelihood approach given  different within-community second-order correlation density $\lambda$ in (\ref{eq:12}). The setting is similar to the weak marginal signal cases. Specifically, the sample networks contain two communities with identical pairwise within-community correlation $\rho = 0.6$. The sizes of the sample networks and nodes are $M=40, N=40$. The density $\lambda$ increases from 0.01 to 1. The Adjusted Rand Index comparisons are illustrated in Figure \ref{fig:2}. In general, the approximate likelihood approach has improving performance when the correlation connectivities among within-community edges increase, in contrast to the independent likelihood approach. Figure \ref{fig:2} shows that the true membership recovery using the  approximate likelihood approach is high even when the second-order within-community correlation is relatively sparse ($\lambda = 0.05$), while the independent likelihood approach performs poorly with a constant ARI regardless of $\lambda$ . This finding supports Theorem \ref{T3.1} and \ref{T3.2} in that the proposed method produces an accelerated decay in misclassification rate as $\lambda$ increases.

\subsection{Study 2: Networks with additional dependence between different communities}
In Study 2, we also investigate whether the proposed method holds for a more general dependency structure among edges from different communities, for example, correlation among edges between different communities
\begin{align}\label{eq:18}
corr(Y^m_{i_1j_1},Y^m_{i_2j_2}) = \tilde{\rho},\;\text{given}\; z_{i_1}=z_{j_1} = q, \;z_{i_2} = z_{j_2}=l,\, q \neq l,
\end{align}
where $\tilde{\rho} \leq \rho_q$ in (\ref{eq:5}) in general. While (\ref{eq:5}) characterizes the concordance of edges within a community, (\ref{eq:18}) also captures the heterogeneity of sample networks. The heterogeneity of multi-layer networks is common in community detection. 

In this simulation, we demonstrate that the proposed method is still robust when there is heterogeneity of connectivities among sample networks. 
To simulate the dependency among inter-community connectivity, we split $M$ sample networks into 10 groups. Within each group, we add the random effects $\gamma_k$ to the within-community marginal means:
\begin{align*}
&\mu_{qq}^m = \frac{exp(\beta_{ql}x^m_{ij})}{1+exp(\beta_{ql}x^m_{ij})}+\gamma_{k},\;M\frac{k-1}{10}\leq m \leq M\frac{k}{10},
\end{align*}
where $\gamma_{k} \sim N(0,\sigma^2)$, $k=1,\cdots,10$, $m =1,\cdots,M$, and $q=1,\dots,K$. The variance $\sigma$ of the random effect $\gamma_k$ captures the intensity of dependency among inter-community connectivities, which increases as $\sigma$ increases. We set $\sigma = 0.5$ to represent a weak inter-community dependency and $\sigma = 1.5$ for a strong inter-community dependency, while the other settings remain the same as in simulation Study 1. Our primary interest is to compare clustering performance between the proposed method and the variational method under the weak marginal signal case. 

Tables \ref{tab:3} and \ref{tab:4} illustrate the clustering performance between the variational method and the proposed method under balanced and unbalanced community sizes, respectively. When the within-community correlation is moderate at 0.3, the proposed method improves the clustering accuracy by $170\%$ to $257\%$ for various network sizes and $\sigma$. For strong correlation $\rho = 0.6$, the improvement is between $210\%$ to $257\%$. In particular, the proposed method has better performance when the networks have strong intra-community correlation and large sample sizes under both weak and strong inter-community correlation cases. In addition, using the fourth-order Bahadur representation improves the accuracy by $6\%$ and $14\%$ when $\sigma = 0.5$ and $\sigma = 1.5$ compared to the second-order Bahadur representation, indicating that the higher-order method still enhances the clustering outcome under the misspecified model. It is interesting to note that the performance of the proposed method decreases by $5\%$ to $15\%$ when the inter-community correlation is strong and the number of networks is small, compared to the same setting with weak inter-community correlation. However, the performances under both weak or strong inter-community correlation are similar when the sample size of networks increases. In conclusion, the proposed method is robust against misspecified dependency structure when the sample size increases.

\section{Real Data Example}
In this section, we apply the proposed method to the 2010 Worldwide Food Import/Export Network dataset \cite{de2015structural} from the Food and Agriculture Organization of the United Nations (\url{http://www.fao.org}). We create 364 networks among 214 countries with a total of 318,346 edges, where each network captures the trading connections of a specific food product among countries.

The primary goal of the study is to identify the food and agricultural product trading network communities among different countries. One significant feature of these networks is that the average empirical correlation of the pairwise connection among trading countries is 0.29. Therefore, the SBM based on the conditional independent assumption among edges could possibly lead to a biased network clustering of countries. 

We first preprocess the data to select nodes corresponding to the trading countries which are most relevant, the number of communities and the initial memberships of countries. Note that several major countries dominate the world economy and lead a high number of trading connectivities, while the other countries with limited agricultural product categories have fewer trading connections with other countries for specific product networks. Here we focus on the partial trading networks consisting of major countries whose corresponding degrees of nodes are larger than 9, which results in 51 countries with major economic impact in the world, such as the United States, mainland China, Japan and some European countries. The average empirical correlation of the trading connections among these countries is 0.22, indicating that the connectivity dependency should be considered in clustering these countries' trading networks. 

In general, there are two major procedures to select the number of communities. First, we can perform the Louvain method for community detection on each individual trading network to obtain the number of communities which maximizes the modularity and the size of the largest community. Next we take the average of the number of communities on networks whose number of communities is smaller than 10 and whose largest community size is larger than 14. This procedure removes the $18\%$ of the product trading networks whose countries are commercially isolated from other countries, as our goal is to detect the commercial communities among the countries which are more connected with other countries. After preprocessing, the average number of communities is 4.9 and we set it to be 4, and there are 296 sample networks remaining in the following analysis.

\begin{figure}[H]
\centering
   \includegraphics[width=5.5in, height=3in]{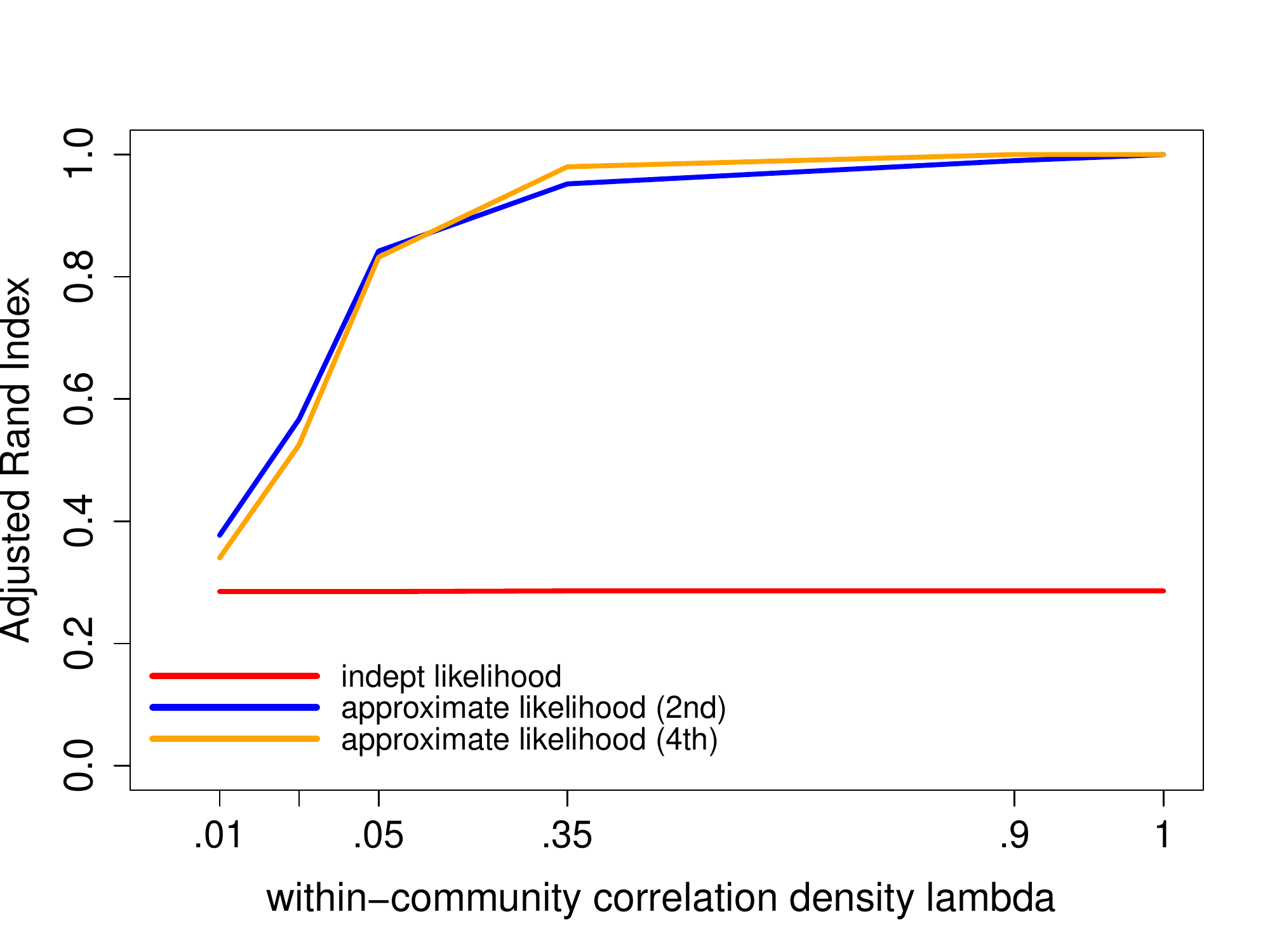}
   \vskip -0.5ex
  \caption{Clustering performance comparisons between independent likelihood and the proposed approximate likelihood approach incorporating the second-order and fourth-order correlations.}
  \label{fig:2}
  \end{figure}

Table \ref{tab:5} and Figure \ref{fig:2} provide the estimated agricultural products trading communities among 51 countries based on the variational EM and the proposed method. For the proposed method, we implement the fourth-order Bahadur approximation since it can better capture high-order within-community connectivity dependency. Table \ref{tab:5} 
presents the clustering outcome among countries according to the variational method and the proposed method. The countries in the same community under the variational method are marked with the same color, while the newly formed communities based on the proposed method are illustrated on the right sides of Table \ref{tab:5} 
and Figure \ref{fig:2}. 
In general, the Adjusted Rand Index for clustering between the variational method and the proposed method is 0.43, indicating that the communities detected by the two methods are quite different. The clustering results from the proposed method incorporating within-community dependency are more interpretable compared to the variational EM using only marginal information.

In particular, the proposed method identifies communities 1 and 2 (red and cyan color communities on the right panel of Figure \ref{fig:2}) which are highly associated with their geographical and climate environments. However, these features are not detected by the variational method. For example, community 1 with the cyan color on the left of Figure \ref{fig:2} 
based on the variational method mainly consists of two types of countries: one group comprises Nordic and Eastern European countries, and the other group consists of countries in Latin American and Africa. In contrast, the proposed method clusters countries from geographically neighboring countries in east Europe, including Austria, Poland and Romania which are clustered with other communities by the variational method. Community 2 with blue color on the left of Figure \ref{fig:2} 
based on the variational method contains northern countries such as Canada as well as tropical countries. However, the proposed method identifies community 2 with tropical coastal countries and Arabian Peninsula countries, which provides more meaningful community clusters compared to the variational EM method.  

The variational method and proposed method detect the same third community with orange color in Figure \ref{fig:2} 
which contains 7 major countries from the European Union: Belgium, France, Germany, Italy, Netherlands, Spain and the UK.

The fourth community from the variational method colored with red on the left of Figure \ref{fig:2} 
consists of 11 Eastern European countries, and all are categorized in community 1 from the proposed method. Community 4 with blue color on the right of Figure \ref{fig:2} 
in the proposed method includes countries with large populations or more developed agricultural product trading, such as mainland China, U.S.A, India and Japan.

In terms of parameter estimation, the average probability of having trading connections for communities 1 and 2 based on the variational method are 0.21 and 0.52, respectively. For the proposed method, the estimated correlations of connectivities within communities 1 and 2 are both 0.22, and the corresponding average within-communities connection rates are 0.28 and 0.22, respectively. The relatively low connection rates and correlations may be related to the low diversity and high overlaps of product categories due to more restrictive geographical and climate environments. 

For community 3, the corresponding estimated marginal parameters $\beta_{33}$ from the proposed method and the variational method are 2.58 and 2.00 respectively, both of which indicate that the trading connection rate within European Union communities is greater than $88\%$ on average. This strong marginal signal of within-community connection explains that the additional correlation information is less influential in clustering. Additionally, the estimated correlation within the third community is 0.58, implying a high connection rate within-community. For community 4, the corresponding average connection rate is 0.49 based on the variational method, and the estimated within-community average connection rate and the correlation are 0.61 and 0.27, respectively. This is because community 4 involves large population countries with more frequent trading on product categories due to their higher food diversity than other countries.

\section{Discussion} 
In this paper, we propose a new community detection method for networks incorporating the underlying dependency structure among connectivities. To model the correlation without specifying a joint likelihood for correlated edges, we construct an approximate likelihood based on the Bahadur representation which decomposes a joint distribution into a marginal term and high-order interaction terms. The proposed method provides  flexible modeling on the correlation structure which can be specified through the interaction term in the approximate likelihood.

In theory, we establish the consistency of the nodes' membership estimator based on the proposed approximate likelihood and show that it achieves a faster convergence than the independent method. In addition, we show that the proposed iterative algorithm possesses desirable convergence properties. In particular, we show that the proposed approximate approach can achieve a faster computational convergence and a lower clustering bias compared to the variational EM algorithm. Furthermore, we show that the variational EM algorithm is a special case of our algorithm under the conditional independent model, which confirms that incorporating correlation information improves the accuracy for community detection.

Our numeric studies indicate that incorporating the within-community correlation among edges can improve the clustering performance compared to the marginal model, even under a moderately misspecified model on inter-community dependency. The improvement of community detection is more significant when the marginal signal is weak, which is less informative for distinguishing between within-community and between-community networks. In addition, the proposed method enables us to achieve more accurate parameter estimation.

In this paper, we only consider incorporating the within-community dependency. It would be worthy of further research to investigate more generalized dependency structures to include between-community dependency as well.

\vspace{-2mm}

\begin{table}[H]

\centering
\caption{Adjusted Rand Index between estimated membership and true membership for networks with two communities and weak marginal signal averaging on 50 replicates.}
\begin{tabular}{c c c c c c c c}
\hline\hline
\multirow{3}{*}{\textbf{}} & &\multicolumn{3}{c}{\textbf{Unbalanced community}}  & \multicolumn{3}{c}{\textbf{Balanced community}}\\
&  &$M = 20$ & $M=40$ & $M=60$ & $M=20$ & $M=40$ & $M=60$ 
\\ \cline{1-4} \cline{5-8}

\multirow{3}{*}{$\rho=0$} &VEM  &0.38& 0.41 & 0.48 & 0.31& 0.28& 0.28  \\
&Bahadur\textsubscript{2nd}& 0.36& 0.41& 0.47& 0.32& 0.29& 0.29\\
&Bahadur\textsubscript{4th}& 0.35 & 0.37 & 0.47 & 0.30 & 0.29 & 0.30 \\
\cline{1-8}

\multirow{3}{*}{$\rho=0.3$} &VEM  & 0.34& 0.34 & 0.34 &0.28 &0.28  &0.28   \\
&Bahadur\textsubscript{2nd}& 0.94 & 0.98 & 0.99 & 0.96 & 0.99 & 1.00  \\
&Bahadur\textsubscript{4th}& 0.96 & 0.99 & 1.00 & 0.99 & 0.99 &1.00  \\
\cline{1-8}

\multirow{3}{*}{$\rho=0.6$} &VEM  &0.34 & 0.34 & 0.34 & 0.29 & 0.28  & 0.28  \\
&Bahadur\textsubscript{2nd}& 0.96 & 0.99 & 0.99 & 0.97 & 1.00 & 1.00  \\
&Bahadur\textsubscript{4th}& 0.99 & 1.00 & 1.00 & 0.99 & 1.00 & 1.00  \\
\hline\hline

\hline

\end{tabular}
\label{tab:1}
\end{table}
\begin{table}[H]

\centering
\caption{Adjusted Rand Index between estimated membership and true membership for networks with two communities and strong marginal signal averaging on 50 replicates.}
\begin{tabular}{c c c c c c c c}
\hline\hline
\multirow{3}{*}{\textbf{}} & &\multicolumn{3}{c}{\textbf{Unbalanced community}}  & \multicolumn{3}{c}{\textbf{Balanced community}}\\
&  &$M = 20$ & $M=40$ & $M=60$ & $M=20$ & $M=40$ & $M=60$ 
\\ \cline{1-4} \cline{5-8}

\multirow{3}{*}{$\rho=0$} &VEM  &0.78& 0.92 & 0.98 & 0.76& 0.90& 0.97  \\
&Bahadur\textsubscript{2nd}& 0.73& 0.91& 0.97& 0.77& 0.92& 0.98\\
&Bahadur\textsubscript{4th}& 0.69 & 0.86 & 0.95 & 0.72 & 0.92 & 0.98 \\
\cline{1-8}

\multirow{3}{*}{$\rho=0.3$} &VEM  & 0.78& 0.81 & 0.83 & 0.68 & 0.79 & 0.84  \\
&Bahadur\textsubscript{2nd}& 0.99 & 0.99 & 1.00 & 0.98 & 1.00 & 1.00  \\
&Bahadur\textsubscript{4th}& 0.99 & 0.99 & 1.00 & 0.99 & 1.00 & 1.00  \\
\cline{1-8}

\multirow{3}{*}{$\rho=0.6$} &VEM  &0.78 & 0.89 & 0.83 & 0.84 & 0.92 & 0.88  \\
&Bahadur\textsubscript{2nd}& 0.99 & 1.00 & 1.00 & 0.99 & 1.00 & 1.00 \\
&Bahadur\textsubscript{4th}& 0.99 & 1.00 & 1.00 & 0.99 & 1.00 & 1.00 \\
\hline\hline

\hline

\end{tabular}
\label{tab:2}
\end{table}
\begin{table}[H]

\centering
\caption{Estimation of within-community parameter $\beta_{11} = 1$ for networks with two communities and weak marginal signal.}
\begin{tabular}{c c c c c c c c}
\hline\hline
\multirow{3}{*}{\textbf{}} & &\multicolumn{3}{c}{\textbf{Unbalanced community}}  & \multicolumn{3}{c}{\textbf{Balanced community}}\\
&  &$M = 20$ & $M=40$ & $M=60$ & $M=20$ & $M=40$ & $M=60$ 
\\ \cline{1-4} \cline{5-8}

\multirow{3}{*}{$\rho=0$} &VEM &$0.56_{0.42}$ & $0.59_{0.29}$ & $0.58_{0.20}$ & $0.64_{0.32}$ & $0.57_{0.16}$ & $0.64_{0.18}$  \\
&Bahadur\textsubscript{2nd}&$0.57_{0.42}$ & $0.58_{0.30}$ & $0.57_{0.21}$ & $0.61_{0.28}$ & $0.57_{0.16}$ & $0.66_{0.20}$\\
&Bahadur\textsubscript{4th}& $0.52_{0.42}$ & $0.55_{0.28}$ & $0.57_{0.19}$  & $0.58_{0.27}$ & $0.58_{0.18}$ & $0.65_{0.19}$ \\
\cline{1-8}

\multirow{3}{*}{$\rho=0.3$} &VEM & $0.49_{0.30}$ & $0.50_{0.17}$ & $0.52_{0.14}$  &  $0.58_{0.24}$ & $0.58_{0.18}$ & $0.59_{0.12}$  \\
&Bahadur\textsubscript{2nd}& $0.81_{0.48}$ & $0.84_{0.32}$ & $0.89_{0.27}$ &  $0.95_{0.24}$ & $0.93_{0.16}$ & $0.92_{0.14}$ \\
&Bahadur\textsubscript{4th}& $0.85_{0.47}$ & $0.83_{0.31}$ & $0.89_{0.27}$ &  $0.96_{0.24}$ & $0.93_{0.16}$ & $0.93_{0.14}$  \\
\cline{1-8}

\multirow{3}{*}{$\rho=0.6$} &VEM & $0.56_{0.22}$ & $0.54_{0.20}$ & $0.52_{0.15}$ &  $0.61_{0.27}$ & $0.61_{0.16}$ & $0.60_{0.14}$ \\
&Bahadur\textsubscript{2nd}& $1.01_{0.42}$ & $1.04_{0.35}$ & $1.00_{0.29}$ &  $0.95_{0.31}$ & $1.00_{0.19}$ & $0.96_{0.15}$  \\
&Bahadur\textsubscript{4th}& $0.99_{0.25}$ & $1.05_{0.15}$ & $1.01_{0.13}$ &  $0.97_{0.31}$ & $1.01_{0.19}$ & $0.97_{0.16}$  \\
\hline\hline

\hline

\end{tabular}
\label{tab:6}
\end{table}
\begin{table}[H]

\centering
\caption{Estimation of within-community parameter $\beta_{22} = 1.5$ for networks with two communities and weak marginal signal.}
\begin{tabular}{c c c c c c c c}
\hline\hline
\multirow{3}{*}{\textbf{}} & &\multicolumn{3}{c}{\textbf{Unbalanced community}}  & \multicolumn{3}{c}{\textbf{Balanced community}}\\
&  &$M = 20$ & $M=40$ & $M=60$ & $M=20$ & $M=40$ & $M=60$ 
\\ \cline{1-4} \cline{5-8}

\multirow{3}{*}{$\rho=0$} &VEM  &$1.43_{0.43}$ & $1.42_{0.34}$ & $1.45_{0.26}$  & $1.18_{0.40}$ & $0.94_{0.16}$ & $0.94_{0.15}$   \\
&Bahadur\textsubscript{2nd}& $1.50_{0.39}$ & $1.49_{0.31}$ & $1.45_{0.25}$ & $1.21_{0.42}$ & $0.93_{0.21}$ & $0.97_{0.22}$\\
&Bahadur\textsubscript{4th}& $1.56_{0.37}$ & $1.49_{0.30}$ & $1.46_{0.23}$  & $1.19_{0.47}$ & $0.94_{0.24}$ & $0.96_{0.22}$ \\
\cline{1-8}

\multirow{3}{*}{$\rho=0.3$} &VEM  & $1.31_{0.23}$ & $1.40_{0.11}$ & $1.37_{0.11}$ & $1.05_{0.21}$ & $0.92_{0.16}$ & $0.92_{0.16}$  \\
&Bahadur\textsubscript{2nd}& $1.56_{0.19}$ & $1.50_{0.10}$ & $1.49_{0.09}$ & $1.48_{0.22}$ & $1.45_{0.19}$ & $1.44_{0.14}$ \\
&Bahadur\textsubscript{4th}& $1.55_{0.19}$ & $1.50_{0.09}$ & $1.49_{0.09}$ & $1.48_{0.22}$ & $1.45_{0.19}$ & $1.45_{0.14}$ \\
\cline{1-8}

\multirow{3}{*}{$\rho=0.6$} &VEM & $1.46_{0.16}$ & $1.43_{0.16}$ & $1.38_{0.13}$ & $1.16_{0.21}$ & $1.09_{0.21}$ & $1.06_{0.22}$ \\
&Bahadur\textsubscript{2nd}& $1.73_{0.29}$ & $1.60_{0.15}$ & $1.52_{0.12}$ & $1.73_{0.28}$ & $1.60_{0.29}$ & $1.64_{0.15}$ \\
&Bahadur\textsubscript{4th}& $1.69_{0.25}$ & $1.60_{0.15}$ & $1.52_{0.13}$ & $1.73_{0.26}$ & $1.61_{0.29}$ & $1.64_{0.15}$  \\
\hline\hline

\hline

\end{tabular}
\label{tab:7}
\end{table}
\begin{table}[H]

\centering
\caption{Estimation of within-community parameter $\beta_{12} = 0$ for networks with two communities and weak marginal signal.}
\begin{tabular}{c c c c c c c c}
\hline\hline
\multirow{3}{*}{\textbf{}} & &\multicolumn{3}{c}{\textbf{Unbalanced community}}  & \multicolumn{3}{c}{\textbf{Balanced community}}\\
&  &$M = 20$ & $M=40$ & $M=60$ & $M=20$ & $M=40$ & $M=60$ 
\\ \cline{1-4} \cline{5-8}

\multirow{3}{*}{$\rho=0$} &VEM  &$0.52_{0.35}$ & $0.57_{0.24}$ & $0.47_{0.22}$  & $0.22_{0.31}$ & $0.39_{0.14}$ & $0.41_{0.11}$  \\
&Bahadur\textsubscript{2nd}& $0.51_{0.32}$ & $0.58_{0.23}$ & $0.48_{0.21}$ & $0.23_{0.30}$ & $0.41_{0.16}$ & $0.39_{0.15}$\\
&Bahadur\textsubscript{4th}& $0.51_{0.29}$ & $0.63_{0.22}$ & $0.48_{0.20}$  & $0.25_{0.28}$ & $0.40_{0.17}$ & $0.41_{0.13}$ \\
\cline{1-8}

\multirow{3}{*}{$\rho=0.3$} &VEM  & $0.68_{0.24}$ & $0.68_{0.13}$ & $0.69_{0.10}$ & $0.42_{0.14}$ & $0.35_{0.12}$ & $0.40_{0.10}$  \\
&Bahadur\textsubscript{2nd}& $-0.02_{0.25}$ & $0.00_{0.15}$ & $0.00_{0.11}$ & $0.03_{0.20}$ & $-0.05_{0.16}$ & $-0.02_{0.12}$ \\
&Bahadur\textsubscript{4th}& $-0.02_{0.24}$ & $0.00_{0.14}$ & $0.00_{0.11}$ & $0.03_{0.18}$ & $-0.06_{0.16}$ & $0.03_{0.12}$ \\
\cline{1-8}

\multirow{3}{*}{$\rho=0.6$} &VEM & $0.72_{0.17}$ & $0.71_{0.11}$ & $0.70_{0.09}$ & $0.41_{0.18}$ & $0.45_{0.11}$ & $0.48_{0.11}$ \\
&Bahadur\textsubscript{2nd}& $-0.05_{0.17}$ & $-0.03_{0.13}$ & $0.02_{0.11}$ & $0.00_{0.19}$ & $0.01_{0.12}$ & $0.03_{0.12}$ \\
&Bahadur\textsubscript{4th}& $-0.04_{0.17}$ & $-0.03_{0.13}$ & $-0.02_{0.11}$ & $-0.02_{0.18}$ & $0.00_{0.12}$ & $0.03_{0.11}$ \\
\hline\hline

\hline

\end{tabular}
\label{tab:8}
\end{table}

\begin{table}[H]

\centering
\caption{Performance comparison given misspecified inter-community correlation with balanced community and weak marginal signal averaging on 50 replicates.}
\begin{tabular}{c c c c c c c c}
\hline\hline
\multirow{3}{*}{\textbf{}} & &\multicolumn{3}{c}{$\sigma = 0.5$}  & \multicolumn{3}{c}{$\sigma = 1.5$}\\
&  &$M = 20$ & $M=40$ & $M=60$ & $M=20$ & $M=40$ & $M=60$ 
\\ \cline{1-4} \cline{5-8}

\multirow{3}{*}{$\rho=0.3$} &VEM  &0.28  &0.28  &0.29  & 0.28 &0.28  & 0.29  \\
&Bahadur\textsubscript{2nd} &0.90  & 0.99 &1.00  & 0.76 &  0.99&0.99\\
&Bahadur\textsubscript{4th} &0.96  & 1.00 &1.00  & 0.87 &  0.98&1.00 \\
\cline{1-8}

\multirow{3}{*}{$\rho=0.6$} &VEM  &0.28  &0.28  & 0.29 &0.28  &0.28  &0.29\\
&Bahadur\textsubscript{2nd} & 0.94 &0.99  &1.00  &0.87  &0.99  &1.00\\
&Bahadur\textsubscript{4th}&0.99  &1.00  &1.00  & 0.94 & 0.99 &1.00 \\
\hline\hline
\hline
\end{tabular}
\label{tab:3}
\end{table}
\begin{table}[H]

\centering
\caption{Performance comparison given misspecified inter-community correlation with unbalanced community and weak marginal signal averaging on 50 replicates.}
\begin{tabular}{c c c c c c c c}
\hline\hline
\multirow{3}{*}{\textbf{}} & &\multicolumn{3}{c}{$\sigma = 0.5$}  & \multicolumn{3}{c}{$\sigma = 1.5$}\\
&  &$M = 20$ & $M=40$ & $M=60$ & $M=20$ & $M=40$ & $M=60$ 
\\ \cline{1-4} \cline{5-8}

\multirow{3}{*}{$\rho=0.3$} &VEM  &0.32  &0.33  &0.33  & 0.33  &0.33  &0.33 \\
&Bahadur\textsubscript{2nd}&0.89  &0.98  & 0.99 & 0.89 & 0.95 &0.97\\
&Bahadur\textsubscript{4th}&0.95  &0.99  &0.99  &0.93  &0.94  &0.94 \\
\cline{1-8}

\multirow{3}{*}{$\rho=0.6$} &VEM &0.34  &0.33  &0.34  &0.33  &0.33  &0.33\\
&Bahadur\textsubscript{2nd}& 0.91 & 0.96 &0.98  & 0.91 &0.95  &0.94\\
&Bahadur\textsubscript{4th}&0.95  &0.96  &0.97  & 0.92 &0.93  &0.92 \\
\hline\hline
\hline
\end{tabular}
\label{tab:4}
\end{table}

\begin{table}[H]
\centering
\caption{Clustering of nations in the agricultural products trading networks given 4 communities}
\begin{tabular}{c c c}
\hline\hline
&\textbf{VEM} & \textbf{Bahadur\textsubscript{4th}}  \\
\multirow{5}{*}{\textbf{Community 1}} &\textcolor{cyan}{Brazil, Denmark, Finland, Ireland}   & \textcolor{red}{Austria}, \textcolor{cyan}{Denmark}, \textcolor{cyan}{Finland}, \textcolor{cyan}{Ireland}, \textcolor{red}{Poland} \\
&\textcolor{cyan}{Lebanon, Russia, Sweden, Switzerland} & \textcolor{cyan}{Russia}, \textcolor{cyan}{Sweden}, \textcolor{cyan}{Switzerland}, \textcolor{cyan}{Turkey}\\
&\textcolor{cyan}{Turkey, Ukraine, Argentina, Israel} &\textcolor{red}{Bulgaria}, \textcolor{red}{Croatia}, \textcolor{red}{Czech}, \textcolor{red}{Greece}, \textcolor{red}{Hungary} \\
&\textcolor{cyan}{Mexico, Norway, Portugal, Chile} &\textcolor{cyan}{Israel},\textcolor{red}{ Lithuania}, \textcolor{cyan}{Norway}, \textcolor{cyan}{Portugal} \\
&\textcolor{cyan}{South Africa, Qatar}   & \textcolor{red}{Romania}, \textcolor{red}{Slovakia}, \textcolor{red}{Slovenia}, \textcolor{cyan}{Ukraine}  \\
 \cline{1-3}
\multirow{4}{*}{\textbf{Community 2}}    
                  &\textcolor{blue}{Australia, Canada, Hong Kong, Mainland} &\textcolor{cyan}{Brazil}, \textcolor{blue}{Hong Kong}, \textcolor{blue}{Taiwan}, \textcolor{blue}{Indonesia}   \\
                  &\textcolor{blue}{Taiwan, India, Indonesia, Malaysia} &\textcolor{cyan}{Lebanon}, \textcolor{blue}{Philippines}, \textcolor{blue}{Korea}, \textcolor{cyan}{Argentina}  \\
                  &\textcolor{blue}{Japan, Philippines, Korea, Singapore} &\textcolor{cyan}{Mexico}, \textcolor{cyan}{Chile}, \textcolor{blue}{New Zealand}  \\
                  & \textcolor{blue}{Thailand, U.S.A, New Zealand} &\textcolor{cyan}{South Africa}, \textcolor{cyan}{Qatar} \\

 \cline{1-3}
\multirow{2}{*}{\textbf{Community 3}}&\textcolor{YellowOrange}{Belgium, France, Germany, Italy}  &\textcolor{YellowOrange}{Belgium, France, Germany, Italy}    \\
                  &\textcolor{YellowOrange}{Netherlands, Spain, United Kingdom} &\textcolor{YellowOrange}{Netherlands, Spain, United Kingdom} \\
                                 
 \cline{1-3}
\multirow{3}{*}{\textbf{Community 4}} &\textcolor{red}{Austria, Poland, Bulgaria, Croatia}  &\textcolor{blue}{Australia,} \textcolor{blue}{Canada}, \textcolor{blue}{Mainland}, \textcolor{blue}{India}   \\
                  & \textcolor{red}{Czech, Greece, Hungary, Lithuania} &\textcolor{blue}{Japan}, \textcolor{blue}{Malaysia}, \textcolor{blue}{Singapore}  \\
                & \textcolor{red}{Romania, Slovakia, Slovenia} &\textcolor{blue}{Thailand}, \textcolor{blue}{U.S.A}  \\

                  \hline\hline

\hline
\end{tabular}
\label{tab:5}
\end{table}
\vspace{-3mm}

\appendix

\section*{APPENDIX: NOTATION AND PROOFS}\label{app}

\subsection{Notation}

In the following, we denote the membership of node as random variable $z_i, i = 1,\cdots, N$. Then $\bm{Z} = \{z_1, z_2, \cdots, z_N\}$. Accordingly, we define the true membership of nodes as $z^*_i \in \{1,2,\cdots,K\}, i = 1,\cdots,N$ and $z^* = \{z_1^*, z_2^*,\cdots, z_N^*\}$.
 We denote $P^*(\cdot) = P(\cdot|\bm{Z}=z^*)$ as the conditional probability of observed networks given the true nodes' membership $z^*$. The number of misclassified nodes is denoted as $r$ such that $\|z - z^*\|_0 = r$ for $z\neq z^*$. Define the t-$th$ sample network as $\bm{Y}^t = (Y^t_{ij})_{N\times N}$ and t-$th$ sample network standardized by $\hat{\mu}_{aa}$ as $\bm{\hat{Y}}^{t,a} = (\hat{Y}^{t,a}_{ij})_{N\times N}$ where $\hat{Y}_{ij}^{t,a}  = \frac{Y^t_{ij} - \hat{\mu}_{aa}}{\sqrt{\hat{\mu}_{aa}(1-\hat{\mu}_{aa})}}, a = 1,\cdots, K, \;t= 1,\cdots, M$. We further define the s-$th$ column of $\bm{\hat{Y}}^{t,a}$ as $\hat{Y}^{t,a}_{\cdot s}$.
 
Denote $\bm{\alpha} = (\alpha_1,\cdots,\alpha_{N})$ as the estimated probability of nodes' memberships. Specifically, let $\alpha_{i} = (\alpha_{i1},\cdots ,\alpha_{iK})_{1\times K}$ be the probability of nodes $i$ belonging to each community where $\sum_{q=1}^K\alpha_{iq}=1,\;i = 1,\cdots,N$. For simplicity of notation, if the subscripts indicate the community then $\alpha_{q} = (\alpha_{1q},\cdots ,\alpha_{Nq})_{1\times N}$ represents the probability of each node belonging to community $q$, where $q = 1,\cdots,K$. Similarly, $z_q^* = \{z^*_{1q},z^*_{2q}, \cdots, z^*_{Nq}\}$ is a binary vector indicating nodes whose true membership belongs to community $q, q= 1,\cdots,K$. Let $vec(\cdot)$ stand for the operation of vectorizing a matrix into a column.

\subsection{Lemmas}
The following two lemmas are introduced as the technical steps in the proofs of Theorem \ref{T3.1}, Theorem \ref{T3.2} and Theorem \ref{T4.2}. The proofs of Lemma 1 and Lemma 2 are provided in the supplemental material. 
\begin{lemma}
Consider function $f_1(x) = \sqrt{\big\{ x\log\frac{\mu_{z_iz_j}}{\mu_{z_i^*z_j^*}}+ (1-x)\log\frac{1-\mu_{z_iz_j}}{1-\mu_{z_i^*z_j^*}}\big\}_+}$ and denote
\begin{align*}
X_t^+ = \{f_1(Y^t_{12}), f_1(Y^t_{13}), \cdots, f_1(Y^t_{N-1,N})\}
\end{align*}
where $\{Y_{ij}^t\}_{N\times N}$ are generated through the stochastic block model in section 3.1 and satisfy condition C1, C2 and C3. Define the covariance matrix of $X_t^+$ as $\Sigma_1$. Then $X_t^+$ is a subgaussian vector, i.e.,
\begin{align*}
L = inf\{\alpha\geq 0: E(\exp(\langle z, X_t^+ -E(X_t^+)\rangle))\leq \exp\{\alpha^2 \langle \Sigma_1z,\;\;z\rangle\}/2, z\in R^{N(N-1)/2}\} \leq C
\end{align*}
for some positive constant $C$.
\end{lemma}

\begin{lemma}
Assume $\{Y_{ij}^t\}_{N\times N}$ are generated through the stochastic block model in section 3.1 and satisfy condition C1, C2 and C4. Then given $M > \mathcal{O}(\frac{1}{\lambda})$, we have $$\bm{P}\Big(\frac{1}{M}\sum_{t=1}^M\sum_{\substack{i<j;k<g\\(i,j)\neq(k,g)}}^N \alpha_{iq}\alpha_{jq}\alpha_{kq}\alpha_{gq}\hat{Y}^{t,q}_{ij}\hat{Y}^{t,q}_{kg} > 0\Big)=1$$ 
as $M, N$ increase for $q = 1,2,\cdots, K$. 
\end{lemma}

\subsection{Proof of Theorem 5.1}

Given the independent model in (\ref{eq:4}), we can simplify the likelihood ratio between a random membership $z$ and the true membership $z^*$ as
\begin{align}\label{ap:1}
\log \frac{P_{ind}(\bm{Y|Z = z})}{P_{ind}(\bm{Y|Z = z^*})} = \frac{1}{M}\sum_{t=1}^M\sum_{i<j}\big\{Y_{ij}^t\log\frac{\mu_{z_iz_j}}{\mu_{z_i^*z_j^*}} + (1-Y_{ij}^t)\log\frac{1-\mu_{z_iz_j}}{1-\mu_{z_i^*z_j^*}} \big\}.
\end{align}
We define two transformation functions $f_1(x)$ and $f_2(x)$ as:
\begin{align*}
f_1(x) = \sqrt{\big\{ x\log\frac{\mu_{z_iz_j}}{\mu_{z_i^*z_j^*}}+ (1-x)\log\frac{1-\mu_{z_iz_j}}{1-\mu_{z_i^*z_j^*}}\big\}_+}, \\
f_2(x) = \sqrt{\big\{ x\log\frac{\mu_{z_iz_j}}{\mu_{z_i^*z_j^*}}+ (1-x)\log\frac{1-\mu_{z_iz_j}}{1-\mu_{z_i^*z_j^*}}\big\}_-}.
\end{align*} 
where $\{\}_{+}$ and $\{\}_{-}$ are positive part and negative part of a random variable.
The previous summation can be decomposed as positive part and negative part: 
$$\log \frac{P_{ind}(\bm{Y|Z = z})}{P_{ind}(\bm{Y|Z = z^*})} = \frac{1}{M}\sum_{t=1}^M\sum_{i<j}\{f_1^2(Y_{ij}^t) - f_2^2(Y_{ij}^t)\}.$$ 
Define the vectorized edges in the $t$ th sample network as:
\begin{align}\label{ap:2}
X_t^+ = \{f_1(Y^t_{12}), f_1(Y^t_{13}), \cdots, f_1(Y^t_{N-1,N})\},
X_t^- = \{f_2(Y^t_{12}), f_2(Y^t_{13}), \cdots, f_2(Y^t_{N-1,N})\}.
\end{align}
Note that each element in $X_t^+$ or $X_t^-$ is a bounded binary random variable. In addition, as $f_1(Y_{ij}^t)$ or $f_2(Y_{ij}^t)$ only rescale $Y_{ij}^t$ then they preserve the within-community correlation among $Y_{ij}^t$. 
Then we consider the following quadratic forms
\begin{align*}
Q_1 = \sum_{t=1}^M\langle X_t^+, X_t^+ \rangle, Q_2 = \sum_{t=1}^M\langle X_t^-, X_t^-  \rangle.
\end{align*}
such that $$\log \frac{P_{ind}(\bm{Y|Z = z})}{P_{ind}(\bm{Y|Z = z^*})} = \frac{1}{M}(Q_1 - Q_2)\;\; \text{and}\;\; E(\log \frac{P_{ind}(\bm{Y|Z = z})}{P_{ind}(\bm{Y|Z = z^*})}) = \frac{1}{M}(EQ_1 - EQ_2).$$ For any $t>0$, we have
\begin{align}\label{ap:6}
P^*\Big\{\frac{P_{ind}(\bm{Y|Z = z})}{P_{ind}(\bm{Y|Z = z^*})}>t\Big\} =  P^*\Big\{(Q_1-EQ_1)-(Q_2-EQ_2)>M(\log t) - E(Q_1-Q_2)\Big\} \nonumber \\
\leq\!\!P^*\Big\{ \!Q_1\!-\!EQ_1 \!>\! \frac{M\log t\! -\! E(Q_1\!-\!Q_2)}{2}\!\Big\}\!+\!P^*\Big\{\!Q_2\!-\!EQ_2\!<\!-\frac{M\log t\!-\!E(Q_1\!-\!Q_2)}{2}\!\Big\} \nonumber \\
 \;\;\;\;\;\leq\!\!\frac{1}{2} P^*\Big\{\!|Q_1\!-\!EQ_1|\!\!>\!\!\frac{M\log t\! -\! E(Q_1\!-\!Q_2)}{2}\!\Big\}\!\! +\!\! \frac{1}{2} P^*\Big\{\!|Q_2\!-\!EQ_2|\!\!>\!\!\frac{M\log t\! -\! E(Q_1\!-\!Q_2)}{2}\!\Big\}.
\end{align}
Next, we estimate each of the term in (\ref{ap:6}).
Given the $\{Y_{ij}^t\}_{t=1}^M$ are binary random variables and the setting that any two within-community edges $Y_{i_1j_1}$ and $Y_{i_2j_2}$ have a nonnegative correlation $corr(Y_{i_1j_1}, Y_{i_2j_2})\geq 0$. Notice that

\begin{align*}
corr\big(f_1(Y_{i_1j_1}), f_1(Y_{i_2j_2})\big) = 
\begin{cases*}
      \;\;corr(Y_{i_1j_1}, Y_{i_2j_2})\;\; \text{if} \;\;\mu_{z_iz_j} \geq \mu_{z_i^*z_j^*}  \\
      -corr(Y_{i_1j_1}, Y_{i_2j_2})\;\; \text{if} \;\;\mu_{z_iz_j} < \mu_{z_i^*z_j^*}     
    \end{cases*}.
\end{align*}
We denote the covariance matrix of $X_t^+$ and $X_t^-$ as $\Sigma_1$ and $\Sigma_2$. Notice that a term in (\ref{ap:1}) is zero only when its corresponding node membership is misclassified. Define the the number of nonzero term in (1) as $N_r$ given $\|z-z^*\|_0=r$. Then we have $N_r = \frac{1}{2}rNM$. According to Lemma 1, $X_t^+$ is a subgaussian vector with a bounded subgaussian norm $L\leq C_1$ where $C_1$ is a positive constant and 
\begin{align}\label{ap:15}
L = inf\{\alpha\geq 0: E(\exp(\langle z, X_t^+ -E(X_t^+)\rangle))\leq \exp\{\alpha^2 \langle \Sigma_1z,z\rangle\}/2\}.
\end{align}
Next we estimate $\|\Sigma_1\|_{F}, \|\Sigma_1\|_{op}$ and $\|\Sigma_2\|_{F}, \|\Sigma_2\|_{op}$ where $\|\cdot \|_{F}$ is the matrix Frobenius norm and $\|\cdot \|_{op}$ is the matrix spectral norm. Denote $$\Lambda = diag\big(\sqrt{\var{\{(X_t^+)_{12}\}}},\sqrt{\var{\{(X_t^+)_{13}\}}},\cdots, \sqrt{\var{\{(X_t^+)_{N-1,N}\}}}\big).$$
Then $\|\Sigma_1\|_{op} = \|\Lambda R \Lambda\|_{op} \leq C_2\|R\|_{op}$ where $R$ is the correlation matrix of $X_t^+$ and 
$$C_2\leq \max_{1\leq i<j\leq n}\var{\{(X_t^+)_{ij}\}}\leq \frac{1}{2}\max\{\log \frac{c_1}{c_2}, \log\frac{1-c_2}{1-c_1}\}.$$ Denote the largest eigenvalue of $R$ as $\lambda_R$. From the Gershgorin circle theorem, we have
\begin{align*}
\lambda_R \leq 1 + \max_{i=1,\cdots,N(N-1)/2}{\sum _ { j \neq i } \left| R _ { i j } \right|}.
\end{align*}
Denote the number of node in the largest community is $N_k$. Note that the misclassification number of node $\|z - z^* \|_0 = r$ and edgewise correlation density $\lambda$ both affect the sparsity of $R$, we have for each row in $R$:
\begin{align*}
\sum _ { j \neq i } \left| R _ { i j } \right| \leq \rho N_k \min(r, \lambda N_k) \leq \rho \kappa_2 N\min(r, \kappa_2\lambda N),
\end{align*}
where $\rho = \displaystyle\max_{i,j}R_{ij}$. Therefore, we have 
$$\|\Sigma_1\|_{op}\leq C_2\{1+\rho \kappa_2 N\min(r, \kappa_2\lambda N)\}.$$
Similarly we have a same upper bound for $\|\Sigma_2\|_{op}$. 
Notice that the dimension of $R$ is $N_r \times N_r$ and $N_r \leq rN$. In each row of $R$, the number of non-zero elements is less than $1+N_k\min(r, \lambda N_k)$. Therefore, we have $$\|\Sigma_1\|_{F}^2 \leq C_2\rho^2rN\{1+\kappa_2 N\min(r, \kappa_2\lambda N)\}.$$
Then we are able to estimate the upper bound for the first term in (\ref{ap:6}). According to the generalized Hanson-Wright inequality in (\cite{chen2018hanson}), we have:
\begin{align}\label{ap:7}
&\frac{1}{2} P^*\Big\{ |Q_1-EQ_1| > s \Big\} \leq \exp\Big\{-C\min\big(\frac{s^2}{L^4\|\Sigma_1\|_{F}^2\|A\|_{F}^2},\frac{s}{L^2\|\Sigma_1\|_{op}\|A\|_{op}}\big)\Big\}.
\end{align}
where $s = \frac{M\log t - E(Q_1-Q_2)}{2}$, $A = \bm{I}_{M\times M}$ and $L$ is subgaussian norm of $X_t^+$ defined in (\ref{ap:15}). Then we have $L \leq C_1$ and $\|A\|^2_{F} = M, \|A\|_{op} = 1$. To estimate $s$, notice
\begin{align*}
E(Q_1 - Q_2) = E[\sum_{t=1}^M\sum_{i<j}\big\{Y_{ij}^t\log\frac{\mu_{z_iz_j}}{\mu_{z_i^*z_j^*}} + (1-Y_{ij}^t)\log\frac{1-\mu_{z_iz_j}}{1-\mu_{z_i^*z_j^*}}\big\}]\\
 = -M\sum_{i<j}\big\{\mu_{z_i^*z_j^*}\log\frac{\mu_{z_i^*z_j^*}}{\mu_{z_iz_j}} + (1-\mu_{z_i^*z_j^*})\log\frac{1-\mu_{z_i^*z_j^*}}{1-\mu_{z_iz_j}}\},
\end{align*}
where there are total $N_r$ non-zero terms in the summation. We introduce the function $$k(x,y) = x\log(x/y)+(1-x)\log(1-x)/(1-y).$$ 
Notice that $k(x,y)>0$ for every $x,y\in (0,1)$. Then we define:
\begin{align}\label{ap:3}
c^*: = \min\{k(\mu_{ql},\mu_{q'l'})\} >0
\end{align} 
where the minimum are taken over 
$\left\{ \left( ( q , l ) , \left( q ^ { \prime } , l ^ { \prime } \right) \right) | \mu _ { q , l } ^ { * } \neq \mu _ { q ^ { \prime } , l ^ { \prime } } ^ { * } \right\}$.
Combined with $N_r = \frac{1}{2}rNM$, we have $-E(Q_1 - Q_2) > \frac{c^*}{2}rNM$. Then for any fixed $t>0$, $s>\mathcal{O}_N(\frac{c^*}{2}rNM)$.
Therefore, we have
\begin{align*}
&\min\big(\frac{s^2}{L^4\|\Sigma_1\|_{F}^2\|A\|_{F}^2},\frac{s}{L^2\|\Sigma_1\|_{op}\|A\|_{op}}\big)\\\geq &\min\big(\frac{(\frac{c^*}{2}rNM)^2}{C_1^2MC_2\rho^2rN\{1+\kappa_2 N\min(r, \kappa_2\lambda N)\}},\frac{\frac{c^*}{2}rNM}{C_1C_2\{1+\rho \kappa_2 N\min(r, \kappa_2\lambda N)}\big)\\
\geq & C_3\frac{rMN}{1+\rho \kappa_2 N\min(r, \kappa_2\lambda N)}.
\end{align*} 
where $C_3 = \frac{\min(c^*,c^{*2})}{2C_1^2\max\{\log \frac{c_1}{c_2}, \log\frac{1-c_2}{1-c_1}\}}$. Hence for (\ref{ap:7}) we have:
\begin{align*}
&\frac{1}{2} P^*\Big\{ |Q_1-EQ_1| > s \Big\} \leq \exp\Big\{-C^*\frac{rMN}{1+\rho \kappa_2 N\min(r, \kappa_2\lambda N)}\Big\}.
\end{align*} 
where $C^* = \frac{C\min(c^*,c^{*2})}{2C_1^2\max\{\log \frac{c_1}{c_2}, \log\frac{1-c_2}{1-c_1}\}} = \frac{c}{\max\{\log \frac{c_1}{c_2}, \log\frac{1-c_2}{1-c_1}\}}$ and $c: = \frac{C\min(c^*,c^{*2})}{2C_1^2}>0$.
Follow Lemma 1, $X_t^-$ is also subgaussian vector. Then we can obtain a same upper bound for 
$$ \frac{1}{2} P^*\Big\{\!|Q_2\!-\!EQ_2|\!\!>\!\!\frac{M\log t\! -\! E(Q_1\!-\!Q_2)}{2}\!\Big\}$$
in (\ref{ap:6}) through the above procedure. Therefore,
\begin{align*}
P^*\Big\{\frac{P_{ind}(\bm{Y|Z = z})}{P_{ind}(\bm{Y|Z = z^*})}>t\Big\} \leq \exp\Big\{-C^*\frac{rMN}{1+\rho \kappa_2 N\min(r, \kappa_2\lambda N)}\Big\}.
\end{align*}

\subsection{Proof of Theorem 5.2}

We continue use the notations in the previous proof of Theorem 5.1. First decompose the proposed approximate likelihood in two parts:
\begin{align*}
\log \frac{\tilde{L}(\bm{Y|Z = z})}{\tilde{L}(\bm{Y|Z = z^*})} &= \log \frac{P_{ind}(\bm{Y|Z = z})}{P_{ind}(\bm{Y|Z = z^*})}\\ &+ \frac{1}{M}\sum_{t=1}^M\log \frac{1+\sum_{k=1}^K\frac{\rho_{k}}{2}\max\Big\{\displaystyle\sum_{\substack{i<j;u<v\\(i,j)\neq(u,v)}}^Nz_{ik}z_{jk}z_{uk}z_{vk}\hat{Y}_{ij}^{t,k}\hat{Y}_{uv}^{t,k}, 0\Big\}}{1+\sum_{k=1}^K\frac{\rho_{k}}{2}\max\Big\{\displaystyle\sum_{\substack{i<j;u<v\\(i,j)\neq(u,v)}}^Nz^*_{ik}z^*_{jk}z^*_{uk}z^*_{vk}\hat{Y}_{ij}^{t,k}\hat{Y}_{uv}^{t,k}, 0\Big\}}.
\end{align*} 
Follow Lemma 2 without assuming condition C4, $P\Big(\displaystyle\sum_{\substack{i<j;u<v\\(i,j)\neq(u,v)}}^Nz^*_{ik}z^*_{jk}z^*_{uk}z^*_{vk}\hat{Y}_{ij}^{t,k}\hat{Y}_{uv}^{t,k} \geq 0\Big)$ goes to 1 as $M, N$ increase. Based on the mean value theorem, we have for some constant $C_1$ that
\begin{align}\label{ap:8}
\log \frac{1+\sum_{k=1}^K\frac{\rho_{k}}{2}\max\Big\{\displaystyle\sum_{\substack{i<j;u<v\\(i,j)\neq(u,v)}}^Nz_{ik}z_{jk}z_{uk}z_{vk}\hat{Y}_{ij}^{t,k}\hat{Y}_{uv}^{t,k}, 0\Big\}}{1+\sum_{k=1}^K\frac{\rho_{k}}{2}\max\Big\{\displaystyle\sum_{\substack{i<j;u<v\\(i,j)\neq(u,v)}}^Nz^*_{ik}z^*_{jk}z^*_{uk}z^*_{vk}\hat{Y}_{ij}^{t,k}\hat{Y}_{uv}^{t,k}, 0\Big\}} \nonumber \\
= C_1\sum_{k=1}^K\frac{\rho_k}{2}\Big\{\max\Big(\sum_{\substack{i<j;u<v\\(i,j)\neq(u,v)}}^Nz_{ik}z_{jk}z_{uk}z_{vk}\hat{Y}_{ij}^{t,k}\hat{Y}_{uv}^{t,k}, 0\Big) - \max\Big(\sum_{\substack{i<j;u<v\\(i,j)\neq(u,v)}}^Nz_{ik}z_{jk}z_{uk}z_{vk}\hat{Y}_{ij}^{t,k}\hat{Y}_{uv}^{t,k}, 0\Big)     \Big\}\nonumber\\
\leq  C_1\sum_{k=1}^K\frac{\rho_k}{2}\Big\{\sum_{\substack{i<j;u<v\\(i,j)\neq(u,v)}}^N(z_{ik}z_{jk}z_{uk}z_{vk} - z^*_{ik}z^*_{jk}z^*_{uk}z^*_{vk})\hat{Y}_{ij}^{t,k}\hat{Y}_{uv}^{t,k}\Big\}.
\end{align}
Notice in summation (\ref{ap:8}), the terms are non-zero only when $z_{ik}z_{jk}z_{uk}z_{vk} \neq z^*_{ik}z^*_{jk}z^*_{uk}z^*_{vk}$. We denote two node sets 
\begin{align*}
\xi_1 = \{(i,j,u,v)|z_{ik}z_{jk}z_{uk}z_{vk} = 1, z^*_{ik}z^*_{jk}z^*_{uk}z^*_{vk} = 0, k=1,\cdots, K\},\\ 
\xi_2 = \{(i,j,u,v)|z^*_{ik}z^*_{jk}z^*_{uk}z^*_{vk} = 1, z_{ik}z_{jk}z_{uk}z_{vk} = 0 ,k=1,\cdots, K\}.
\end{align*}
where $\#|\xi_1| = N_1$ and $\#|\xi_2| = N_2$.
Given the number of misclassified nodes $\|z - z^* \|_0 = r$, we have $N_1 =  \mathcal{O}(rN^3)$ and $N_2 = \mathcal{O}(rN^3) $. In the following, we construct the augmented edge vectors for the $t$ th sample network by incorporating the vectorized pairwise edge interaction in (\ref{ap:8}) such that:
\begin{align*}
\tilde{X}_t^+ = \Big\{X_t^+, \underbrace{\Big(\sqrt{\frac{C_1\rho_k}{2}\{\hat{Y}_{ij}^{t,k}\hat{Y}_{uv}^{t,k}\}_{+}}\Big)_{1\times N_1}}_{\substack{(i,j,u,v)\in \xi_1\\z_{ik}z_{jk}z_{uk}z_{vk} = 1\\k = 1,\cdots,K}}, \underbrace{\Big(\sqrt{\frac{C_1\rho_k}{2}\{-\hat{Y}_{ij}^{t,k}\hat{Y}_{uv}^{t,k}\}_{+}}\Big)_{1\times N_2}}_{\substack{(i,j,u,v)\in \xi_2\\z^*_{ik}z^*_{jk}z^*_{uk}z^*_{vk} = 1\\k=1,\cdot,K}}  \Big\},\\
\tilde{X}_t^- = \Big\{X_t^-, \underbrace{\Big(\sqrt{\frac{C_1\rho_k}{2}\{\hat{Y}_{ij}^{t,k}\hat{Y}_{uv}^{t,k}\}_{-}}\Big)_{1\times N_1}}_{\substack{(i,j,u,v)\in \xi_1\\z_{ik}z_{jk}z_{uk}z_{vk} = 1\\k=1,\cdots,K}}, \underbrace{\Big(\sqrt{\frac{C_1\rho_k}{2}\{-\hat{Y}_{ij}^{t,k}\hat{Y}_{uv}^{t,k}\}_{-}}\Big)_{1\times N_2}}_{\substack{(i,j,u,v)\in \xi_2\\z^*_{ik}z^*_{jk}z^*_{uk}z^*_{vk} = 1\\k=1,\cdots,K}}  \Big\}.
\end{align*}
where $X_t^+$ and $X_t^-$ are defined in (\ref{ap:2}). Denote the covariance matrix for $\tilde{X}_t^+$ and $\tilde{X}_t^-$ are $\tilde{\Sigma}_1$ and $\tilde{\Sigma}_2$ respectively. Since the second-order terms in $X_t^+$ and $X_t^-$ such as $\sqrt{\frac{C_1\rho_k}{2}\{\hat{Y}_{ij}^{t,k}\hat{Y}_{uv}^{t,k}\}_{+}}$ only rescale the original edgewise interaction $\hat{Y}_{ij}^{t,k}\hat{Y}_{uv}^{t,k}$ then they preserve the third-order and fourth-order correlation within communities such that
$$|E\Big\{f_1(Y_{i_1j_1}^t)\sqrt{\frac{C\rho_k}{2}\{\hat{Y}_{i_2j_2}^{t,k}\hat{Y}_{i_3j_3}^{t,k}\}_{+}}\Big\}| = |E(\hat{Y}_{i_1j_1}^{t,k}\hat{Y}_{i_2j_2}^{t,k}\hat{Y}_{i_3j_3}^{t,k})|,$$
$$|E\Big\{f_2(Y_{i_1j_1}^t)\sqrt{\frac{C\rho_k}{2}\{\hat{Y}_{i_2j_2}^{t,k}\hat{Y}_{i_3j_3}^{t,k}\}_{-}}\Big\}| = |E(\hat{Y}_{i_1j_1}^{t,k}\hat{Y}_{i_2j_2}^{t,k}\hat{Y}_{i_3j_3}^{t,k})|,$$
$$|E\Big\{\sqrt{\frac{C\rho_k}{2}\{\hat{Y}_{i_1j_1}^{t,k}\hat{Y}_{i_2j_2}^{t,k}\}_{+}}\sqrt{\frac{C\rho_k}{2}\{\hat{Y}_{i_3j_3}^{t,k}\hat{Y}_{i_4j_4}^{t,k}\}_{+}}\Big\}| = |E(\hat{Y}_{i_1j_1}^{t,k}\hat{Y}_{i_2j_2}^{t,k}\hat{Y}_{i_3j_3}^{t,k}\hat{Y}_{i_4j_4}^{t,k})|,$$
$$|E\Big\{\sqrt{\frac{C\rho_k}{2}\{\hat{Y}_{i_1j_1}^{t,k}\hat{Y}_{i_2j_2}^{t,k}\}_{-}}\sqrt{\frac{C\rho_k}{2}\{\hat{Y}_{i_3j_3}^{t,k}\hat{Y}_{i_4j_4}^{t,k}\}_{-}}\Big\}| = |E(\hat{Y}_{i_1j_1}^{t,k}\hat{Y}_{i_2j_2}^{t,k}\hat{Y}_{i_3j_3}^{t,k}\hat{Y}_{i_4j_4}^{t,k})|.$$

Notice that each element in $\tilde{X}_t^+$ or $\tilde{X}_t^-$ is a bounded binary random variable. Follow the same procedure in Lemma 1, we can show that both $\tilde{X}_t^+$ and $\tilde{X}_t^-$ are subgaussian random vectors such that $L_1 \leq C_2, L_2 \leq C_2$ for some constant $C_2$ where $L_1, L_2$ are subgaussian norm of $\tilde{X}_t^+$ and $\tilde{X}_t^-$. 
Then consider the following quadratic forms:
\begin{align*}
\tilde{Q}_1 = \sum_{t=1}^M\langle \tilde{X}_t^+, \tilde{X}_t^+ \rangle, \tilde{Q}_2 = \sum_{t=1}^M\langle \tilde{X}_t^-, \tilde{X}_t^-  \rangle.
\end{align*}
Therefore, we have
\begin{align*}
\log \frac{\tilde{L}(\bm{Y|Z = z})}{\tilde{L}(\bm{Y|Z = z^*})} \leq \frac{1}{M}(\tilde{Q}_1 - \tilde{Q}_2).
\end{align*}

Similar to (\ref{ap:6}), for any fixed $t>0$:
\begin{align}\label{ap:9}
&P^*\Big\{\frac{\tilde{L}(\bm{Y|Z = z})}{\tilde{L}(\bm{Y|Z = z^*})}>t\Big\} \leq P^*\Big\{\frac{1}{M}(\tilde{Q}_1 - \tilde{Q}_2)>\log t\Big\}  \nonumber\\
= &P^*\Big\{(\tilde{Q}_1-E\tilde{Q}_1)-(\tilde{Q}_2-E\tilde{Q}_2)>M(\log t) - E(\tilde{Q}_1-\tilde{Q}_2)\Big\} \nonumber \\
\leq &P^*\Big\{ \tilde{Q}_1-E\tilde{Q}_1 > \frac{M\log t - E(\tilde{Q}_1-\tilde{Q}_2)}{2}\Big\} + P^*\Big\{ \tilde{Q}_2-E\tilde{Q}_2 < -\frac{M\log t - E(\tilde{Q}_1-\tilde{Q}_2)}{2}\Big\} \nonumber \\
 = & \frac{1}{2} P^*\Big\{ |\tilde{Q}_1-E\tilde{Q}_1| > \frac{M\log t - E(\tilde{Q}_1-\tilde{Q}_2)}{2}\Big\} + \frac{1}{2} P^*\Big\{ |\tilde{Q}_2-E\tilde{Q}_2| > \frac{M\log t - E(\tilde{Q}_1-\tilde{Q}_2)}{2}\Big\}.
\end{align}
Next we estimate $\|\tilde{\Sigma}_1\|_{F}, \|\tilde{\Sigma_1}\|_{op}$ and $\|\tilde{\Sigma}_2\|_{F}, \|\tilde{\Sigma}_2\|_{op}$. Denote 
\begin{align*}
\tilde{\Lambda} = diag(\Lambda, sd{\underbrace{\Big(\sqrt{\frac{\rho_k}{2}\{\hat{Y}_{ij}^{t,k}\hat{Y}_{uv}^{t,k}\}_{+}}\Big)_{1\times N_1}}_{\substack{(i,j,u,v)\in \xi_1\\z_{ik}z_{jk}z_{uk}z_{vk} = 1\\k=1,\cdots,K}}},sd{\underbrace{\Big(\sqrt{\frac{\rho_k}{2}\{-\hat{Y}_{ij}^{t,k}\hat{Y}_{uv}^{t,k}\}_{+}}\Big)_{1\times N_2}}_{\substack{(i,j,u,v)\in \xi_2\\z^*_{ik}z^*_{jk}z^*_{uk}z^*_{vk} = 1\\k=1,\cdots,K}}}),
\end{align*}
then $\|\tilde{\Sigma}_1\|_{op} = \|\tilde{\Lambda} \tilde{R} \tilde{\Lambda}\|_{op} \leq C_3\|\tilde{R}\|_{op}$ where $\tilde{R}$ is the correlation matrix of $\tilde{X}_t^+$ and $C_3$ is the largest variance of elements in $\tilde{X}_t^+$. Denote the largest eigenvalue of $\tilde{R}$ as $\lambda_{\tilde{R}}$. 
From the Gershgorin circle theorem, we have
\begin{align*}
\lambda_{\tilde{R}} \leq 1 + \max_{i}\sum _ { j \neq i } \left| \tilde{R} _ { i j } \right|.
\end{align*}
Given that the misclassification number of node $\|z - z^* \|_0 = r$, edgewise correlation density $\lambda$ and condition C3, for each row in $\tilde{R}$, there exists some constant $C_4$>0 such that:
\begin{align}\label{ap:4}
\sum _ { j \neq i } \left| R _ { i j } \right| \leq C_4\rho N_k \min(r, \lambda N_k) = C_4\rho \kappa_2 N\min(r, \kappa_2\lambda N),
\end{align}
where $\rho = \displaystyle\max_{i,j}\tilde{R}_{ij}$. Therefore, we have $$\|\tilde{\Sigma}_1\|_{op}\leq C_3\{1+C_4\rho \kappa_2 N\min(r, \kappa_2\lambda N)\}.$$ 
Similarly, $\|\tilde{\Sigma}_2\|_{op}$ follows a same upper bound. 
Notice that the dimension of $\tilde{R}$ is $(N_r + N_1 +N_2) \times (N_r+N_1+N_2)$. Under the condition C3, in each row of $\tilde{R}$, the number of non-zero elements is less than $1+C_4 N_k\min(r, \lambda N_k)$. Therefore, we have for a constant $C'>0$:
\begin{align*}
\|\tilde{\Sigma}_1\|_{F}^2 \leq & C_3\rho^2(N_r+N_1+N_2)\{1+C_4\kappa_2 N\min(r, \kappa_2\lambda N)\}\\
\leq & C'\rho^2(rN+rN^3)\{1+C_4\kappa_2 N\min(r, \kappa_2\lambda N)\}.
\end{align*}
According to the generalized Hanson-Wright inequality in (\cite{chen2018hanson}):
\begin{align}\label{ap:10}
&\frac{1}{2} P^*\Big\{ |\tilde{Q}_1-E\tilde{Q}_1| > s \Big\} \leq \exp\Big\{-C\min\big(\frac{s^2}{L_1^4\|\tilde{\Sigma}_1\|_{F}^2\|A\|_{F}^2},\frac{s}{L_1^2\|\tilde{\Sigma}_1\|_{op}\|A\|_{op}}\big)\Big\},
\end{align}
where $s = \frac{M\log t - E(\tilde{Q}_1-\tilde{Q}_2)}{2}$, $A = \bm{I}_{M\times M}$ and $L_1$ is subgaussian norm of $\tilde{X}_t^+$.
Notice $\|A\|^2_{F} = M, \|A\|_{op} = 1$. Given (\ref{ap:8}), we have 
\begin{align*}
E(\tilde{Q}_1 - \tilde{Q}_2) = E(Q_1 - Q_2) +  C_1\sum_{k=1}^K\frac{\rho_k}{2}\Big\{\sum_{\substack{i<j;u<v\\(i,j)\neq(u,v)}}^N(z_{ik}z_{jk}z_{uk}z_{vk} - z^*_{ik}z^*_{jk}z^*_{uk}z^*_{vk})E(\hat{Y}_{ij}^{t,k}\hat{Y}_{uv}^{t,k})\Big\}.
\end{align*}
Denote $\rho_{min}$ as the lower bound of all non-zero correlation among edges such that $E(\hat{Y}_{ij}^{t,k}\hat{Y}_{uv}^{t,k}) = \rho_{ijuv} \geq \rho_{min}$. Given the edges from different communities are independent and within-community correlation density $\lambda$, 
we have for some positive constant $C_5$,
$$\#|\{(i,j,u,v): E(\hat{Y}_{ij}^{t,k}\hat{Y}_{uv}^{t,k})>0, (i,j,u,v)\in \xi_2\}| = \lambda N_1 = \lambda C_5rN^3,$$ $$\#|\{(i,j,u,v): E(\hat{Y}_{ij}^{t,k}\hat{Y}_{uv}^{t,k})>0, (i,j,u,v)\in \xi_1\}| \leq \lambda {{r}\choose{4}}.$$ Assume that $r \leq c N$ for some constant $0<c<1$, we have for some constant $c_0>0$: 
\begin{align*}
-E(\tilde{Q}_1 - \tilde{Q}_2) \geq \frac{c^*}{2}rNM + \lambda M\frac{C_1\rho_{min}^2}{2}(C_5rN^3 - {{r}\choose{4}})\geq 
c_0r(N + \lambda N^3)M .
\end{align*}
Given any fixed $t>0$, $s>\mathcal{O}_N(c_0r(N + \lambda N^3)M)$. For the first term in (\ref{ap:10}), 
\begin{align*}
\frac{s^2}{L_1^4\|\tilde{\Sigma}_1\|_{F}^2\|A\|_{F}^2} \geq \frac{c_0^2r^2(N + \lambda N^3)^2M^2 }{L_1^4C'\rho^2(rN+rN^3)\{1+C_4\kappa_2 N\min(r, \kappa_2\lambda N)\}M}.
\end{align*}
For the second term in (\ref{ap:10}),
\begin{align*}
\frac{s}{L_1^2\|\tilde{\Sigma}_1\|_{op}\|A\|_{op}} \geq \frac{c_0r(N + \lambda N^3)M}{L_1^2C'\{1+C_4\rho \kappa_2 N\min(r, \kappa_2\lambda N)\}}.
\end{align*}
Given $\lambda > 0$, we have for some constant $C_6>0$
\begin{align}\label{ap:11}
\min\big(\frac{s^2}{L_1^4\|\tilde{\Sigma}_1\|_{F}^2\|A\|_{F}^2},\frac{s}{L_1^2\|\tilde{\Sigma}_1\|_{op}\|A\|_{op}}\big) \geq C_6 \frac{r\lambda NM(1+\lambda N^2)}{1+C_4\rho \kappa_2 N\min(r, \kappa_2\lambda N)}.
\end{align}
Follow the same procedure we can show a upper bound for $P^*\Big\{ |\tilde{Q}_2-E\tilde{Q}_2| > s \Big\}$ with a same order to (\ref{ap:11}). Combined with (\ref{ap:9}) and (\ref{ap:10}), we have for $\lambda >0$ and some constant $C>0$:
\begin{align*}
P _ { Z* }\Big\{ \frac {\tilde{L}(\bm{Y}| \bm{Z}= z; \Theta)} { \tilde{L}(\bm{Y}| \bm{Z }= z^*; \Theta)} > t \Big\} \leq \exp\Big\{-C\frac{r\lambda NM(1+\lambda N^2)}{1+C_4\rho \kappa_2 N\min(r, \kappa_2\lambda N)}\Big\},
\end{align*}

\subsection{Proof of Theorem 5.3}

Follow the notations introduced in Theorem 5.1 and Theorem 5.2, we further define that $\bm{w} = \max{\frac{P^{(s)}(Z_{i}=q)}{P^{(s)}(Z_{i}=l)}},\;i=1,\cdots,N,\;q,l = 1,\cdots,K$. Let $\bm{E}$ stands for the operator of expectation step in Algorithm 1 in Section 4. 

We first consider the misclassification of updated estimated membership for node $s$, e.g., $\bm{E}(z_s)$ from the current estimation $\alpha_s$. We denote that $\bm{\alpha}_{-s}$ as the probability estimations of nodes' memberships at current step except node $s$ and assume the true membership of node $s$ is $b$, i.e., $z^*_{s} = b$. 
If we use the marginal likelihood, then:
\begin{align} \label{ap:12}
&\| \bm{E}(z_{s}) - z^*_s   \|_1 = \nonumber \\& |\frac{P(z_s=1)\tilde{L}(\bm{Y}|\bm{\alpha}_{-s}; z_s=1)}{\sum_{q=1}^K P(z_s=q)\tilde{L}(\bm{Y}|\bm{\alpha}_{-s}; z_s=q)} - 0 | \!\!+\cdot\cdot\cdot +\!\!| \frac{P(z_s=b)\tilde{L}(\bm{Y}|\bm{\alpha}_{-s}; z_s=b)}{\sum_{q=1}^K P(z_s=K)\tilde{L}(\bm{Y}|\bm{\alpha}_{-s}; z_s=K)}\!-\! 1| \nonumber \\ 
&\leq 2 \frac{\sum_{q\neq b} P(z_s=q)\tilde{L}(\bm{Y}|\bm{\alpha}_{-s}; z_s=q)}{\sum_{q=1}^K P(z_s=q)\tilde{L}(\bm{Y}|\bm{\alpha}_{-s}; Z_s=q)} \leq 2\bm{w}\sum_{q\neq b}^K\frac{\tilde{L}(\bm{Y}|\bm{\alpha}_{-s}; z_s=q)}{\tilde{L}(\bm{Y}|\bm{\alpha}_{-s}; z_s=b)} \nonumber \\
&= 2\bm{w}\sum_{q\neq b}^K \min[1, \exp\{\log \tilde{L}(\bm{Y}|\bm{\alpha}_{-s}; Z_s=q)-\log \tilde{L}(\bm{Y}|\bm{\alpha}_{-s}; z_s=b)\}]. 
\end{align} 
Then given node $s$ belongs to different communities while the estimated membership for other nodes $\bm{\alpha}_{-s}$ are fixed. We decompose the proposed approximate likelihood into marginal part and correlation part in the following: 
$\log \tilde{L}(\bm{Y}|\bm{\alpha}_{-s}; z_s) = \log L_{mar}(\bm{Y}|\bm{\alpha}_{-s}; z_s) + \log L_{cor}(\bm{Y}|\bm{\alpha}_{-s}; z_s)$. The marginal likelihood $\log L_{mar}(\bm{Y}|\bm{\alpha}_{-s}; z_s)$,
\begin{align*}
&\log L_{mar}(\bm{Y}|\bm{\alpha}_{-s}; z_s=a)\\ 
&= \frac{1}{M}\sum_{t=1}^M \Bigg[ \underbrace{log \prod_{q,l}^K\prod_{i\neq j\neq s}^N\big\{\mu_{ql}^{Y^t_{ij}}(1-\mu_{ql})^{(1-Y^t_{ij})}\big\}^{\alpha_{iq}\alpha_{jl}}}_{\text{not depend on}\;z_s } + \prod_{q=1}^K\prod_{i\neq s}^N\big\{\mu_{qa}^{Y^t_{is}}(1-\mu_{qa})^{(1-Y^t_{is})}\big\}^{\alpha_{iq}} \Bigg].
\end{align*}
\vspace{-1mm}
Therefore, the discrepancy among marginal likelihood is 
\begin{align*} 
&\log L_{mar}(\bm{Y}|\bm{\alpha}_{-s}; z_s=a) - \log L_{mar}(\bm{Y}|\bm{\alpha}_{-s}; z_s=b)\\
 = & \frac{1}{M}\sum_{t=1}^M\sum_{q=1}^K\sum_{i\neq s}^N\Big[\alpha_{iq}\{Y_{is}^t \log\frac{\hat{\mu}_{qa}}{\hat{\mu}_{qb}} + (1-Y_{is}^t) \log\frac{1-\hat{\mu}_{qa}}{1-\hat{\mu}_{qb}}\}\Big]\\
 = & \frac{1}{M}\sum_{t=1}^M\sum_{q=1}^K\sum_{i\neq s}^N\Big[\alpha_{iq}\{Y_{is}^t \log\frac{\mu_{qa}}{\mu_{qb}} + (1-Y_{is}^t) \log\frac{1-\mu_{qa}}{1-\mu_{qb}}\}\Big]\\
 + &  \frac{1}{M}\sum_{t=1}^M\sum_{q=1}^K\sum_{i\neq s}^N\Big[\alpha_{iq}\{Y_{is}^t \log\frac{\mu_{qa}\hat{\mu}_{qb}}{\hat{\mu}_{qa}\mu_{qb}} + (1-Y_{is}^t) \log\frac{(1-\mu_{qa})(1-\hat{\mu}_{qb})}{(1-\hat{\mu}_{qa})(1-\hat{\mu}_{qb})}\}\Big]\\
\end{align*}
We can decompose the marginal discrepancy into four parts:
\begin{align*} 
&\log L_{mar}(\bm{Y}|\bm{\alpha}_{-s}; z_s=a) - \log L_{mar}(\bm{Y}|\bm{\alpha}_{-s}; z_s=b)\\
= &\underbrace{\frac{1}{M}\sum_{t=1}^M\sum_{q=1}^K\sum_{i\neq s}^N(\alpha_{iq} -z_{iq}^*)\{Y_{is}^t - E(Y_{is}^t)\}(\log\frac{\mu_{qa}}{\mu_{qb}}-\log\frac{1-\mu_{qa}}{1-\mu_{qb}})}_{\bm{A_1}}\\
+ & \underbrace{\frac{1}{M}\sum_{t=1}^M\sum_{q=1}^K\sum_{i\neq s}^N\Big[(\alpha_{iq}-z^*_{iq})\{EY_{is}^t \log\frac{\mu_{qa}}{\mu_{qb}} + (1-EY_{is}^t) \log\frac{1-\mu_{qa}}{1-\mu_{qb}}\}\Big]}_{\bm{A_2}}\\
+ & \underbrace{\frac{1}{M}\sum_{t=1}^M\sum_{q=1}^K\sum_{i\neq s}^N\Big[z^*_{iq}\{Y_{is}^t \log\frac{\mu_{qa}}{\mu_{qb}} + (1-Y_{is}^t) \log\frac{1-\mu_{qa}}{1-\mu_{qb}}\}\Big]}_{\bm{A_3}}\\
 + &  \underbrace{\frac{1}{M}\sum_{t=1}^M\sum_{q=1}^K\sum_{i\neq s}^N\Big[\alpha_{iq}\{Y_{is}^t \log\frac{\mu_{qa}\hat{\mu}_{qb}}{\hat{\mu}_{qa}\mu_{qb}} + (1-Y_{is}^t) \log\frac{(1-\mu_{qa})(1-\hat{\mu}_{qb})}{(1-\hat{\mu}_{qa})(1-\hat{\mu}_{qb})}\}\Big]}_{\bm{A_4}}.\\
\end{align*}
For the correlation part, we consider the pairwise interaction terms in the $\log L_{cor}(\bm{Y}|\bm{\alpha})$. Notice that for $t=1,\cdots,M$
\vspace{-3mm}
$$\sum_{\substack{i<j;k<g\\(i,j)\neq(k,g)}}^N \alpha_{ia}\alpha_{ja}\alpha_{ka}\alpha_{ga}\hat{Y}^{t,a}_{ij}\hat{Y}^{t,a}_{kg} = (\sum_{i\neq s}^N 
\alpha_{sa}\alpha_{ia}\hat{Y}^{t,a}_{si})(\sum_{i<j}^N\alpha_{ia}\alpha_{ja}\hat{Y}^{t,a}_{ij}) - \sum_{i\neq s}^N 
(\alpha_{ia}\hat{Y}^{t,a}_{si})^2  + A_a^t,$$
\vspace{-0.1cm}
where $A_q^t$ does not depend on $z_s$. Since the first term $(\sum_{i\neq s}^N 
\alpha_{sa}\alpha_{ia}\hat{Y}^{t,a}_{si})(\sum_{i<j}^N\alpha_{ia}\alpha_{ja}\hat{Y}^{t,a}_{ij}) = o(N^3)$ and the second term $\sum_{i\neq s}^N 
(\alpha_{ia}\hat{Y}^{t,a}_{si})^2 = o(N)$, without loss of generality, we can keep the first dominating term when $N$ is large. For the correlation part $\log L_{cor}(\bm{Y}|\bm{\alpha}_{-s}; z_s)$, if $\alpha_{sq}=0,\;q \neq a$ and $\alpha_{sa} = 1$, with probability approaching 1 from Lemma 2 as $M$ increases given $M > \mathcal{O}(\frac{1}{\lambda})$:
\vspace{-0.6cm}
\begin{align*}
\log L_{cor}(\bm{Y}|\bm{\alpha}_{-s}; Z_s=a) =& \frac{1}{M}\sum_{t=1}^M\Big\{1+\sum_{q=1}^K \frac{\rho_{q}}{2} \max(\sum_{\substack{i<j;k<g\\(i,j)\neq(k,g)}}^N \alpha_{iq}\alpha_{jq}\alpha_{kq}\alpha_{gq}\hat{Y}^{t,q}_{ij}\hat{Y}^{t,q}_{kg}, 0)\Big\}\\
=& 1+ \underbrace{\frac{1}{M}\sum_{t=1}^M\sum_{q=1}^K\frac{\rho_{q}}{2}A_q^t}_{\bm{A}}+\underbrace{\frac{\rho_a}{2}(\sum_{i\neq s}^N \alpha_{sa}\alpha_{ia}\hat{Y}^{t,a}_{si})(\sum_{i<j}^N\alpha_{ia}\alpha_{ja}\hat{Y}^{t,a}_{ij}) }_{\bm{B}_a}.
\end{align*}
Through the Taylor expansion, the discrepancy of correlation information when node $s$ belongs to different communities $a$ and $b$:
\begin{align*}
\log L_{cor}(\bm{Y}|\bm{\alpha}_{-s}; Z_s=a)-\log L_{cor}(\bm{Y}|\bm{\alpha}_{-s}; Z_s=b)&= \log(1+\bm{A}+\bm{B_a}) - \log(1+\bm{A}+\bm{B_b}) \\ \nonumber 
&=\log(1+\frac{\bm{B_a}-\bm{B_b}}{1+\bm{A}+\bm{B_b}} )
\leq \bm{C}_{A}(\bm{B_a}-\bm{B_b}),
\end{align*}
where $\bm{C}_{A}$ is a constant relating to the gradient of function $\log (1 + 1/x)$ at $\bm{A}$. Then we set $\rho = \min{\rho_q}, q = 1,\cdots,K$
 \begin{align*}
\bm{B_a}-\bm{B_b} &= (\sum_{i\neq s}^N \alpha_{ia}\hat{Y}^{t,a}_{si})(\sum_{i<j}^N\alpha_{ia}\alpha_{ja}\hat{Y}^{t,a}_{ij}) - (\sum_{i\neq s}^N\alpha_{ib}\hat{Y}^{t,b}_{si})(\sum_{i<j}^N\alpha_{ib}\alpha_{jb}\hat{Y}^{t,b}_{ij})\\
&\leq\frac{\rho}{4}\Big(\langle \alpha_a\otimes vec(\alpha_a^T\alpha_a), \hat{Y}^{t,a}_{\cdot s}\otimes vec(\hat{\bm{Y}}^{t,a}) \rangle - \langle   \alpha_b\otimes vec(\alpha_b^T\alpha_b), \hat{Y}^{t,b}_{\cdot s}\otimes vec(\hat{\bm{Y}}^{t,b}) \rangle\Big).
 \end{align*}

For the simplicity of notation, we define and decompose the correlation discrepancy as followings:
\begin{align*}
\bm{B}:=&\sum_{t=1}^M\frac{\rho C_A}{4M}\big(\langle \alpha_a\otimes vec(\alpha_a^T\alpha_a), \hat{Y}^{t,a}_{\cdot s}\otimes vec(\hat{\bm{Y}}^{t,a}) \rangle \!-\! \langle   \alpha_b\otimes vec(\alpha_b^T\alpha_b), \hat{Y}^{t,b}_{\cdot i}\otimes vec(\hat{\bm{Y}}^{t,b}) \rangle\big)\\
&=\underbrace{\frac{\rho C_A}{4M}\sum_{t=1}^M\big(\langle \alpha_a\otimes vec(\alpha^T_a\alpha_a) - z_a^*\otimes vec(z_a^{*T}z_a^*), \hat{Y}^{t,a}_{\cdot s}\otimes vec(\hat{\bm{Y}}^{t,a})\rangle-}_{}\nonumber \\&\underbrace{\langle \alpha_b\otimes vec(\alpha^T_b\alpha_b) - z_b^*\otimes vec(z_b^{*T}z_b^*), \hat{Y}^{t,b}_{\cdot s}\otimes vec(\hat{\bm{Y}}^{t,b})\rangle\big)}_{\text{misclassification error:}\bm{B}_1}\nonumber \\
&+\underbrace{\frac{\rho C_A}{4M}\sum_{t=1}^M\big(\langle z_a^*\otimes vec(z_a^{*T}z_a^*), \hat{Y}^{t,a}_{\cdot s}\otimes vec(\hat{\bm{Y}}^{t,a}) \rangle - \langle   z_b^*\otimes vec(z_b^{*T}z_b^*), \hat{Y}^{t,b}_{\cdot s}\otimes vec(\hat{\bm{Y}}^{t,b}) \rangle\big)}_{\text{estimation bias:}\bm{B}_2}.
\end{align*}
Notice that $\min\{1, \exp(x)\}\leq \exp(x_0) + \sum_{l=0}^{m-1}\frac{1-\exp(x_0)}{m}\mathbbm{1}  \{x\geq (1-l/m)x_0\}$ and set $x_0 = -\alpha' MN$, where $\alpha'>0$. Given (\ref{ap:12}), we have:
\begin{align}
E\|\bm{\alpha}^{s+1} - \bm{z^*}\|_1 \leq 2\bm{w}NK\exp(-\alpha' MN) + 2\bm{w}\sum_{l=0}^{m-1}\sum_{a=1}^K\sum_{b\neq a}\sum_{i:z_i^*= b}\frac{1-exp(\alpha' MN)}{m}E(\bm{L}_2)
\label{ap:13}
\end{align}
where $E(\bm{L}) = \mathbbm{P}\Big(\bm{A}+\bm{B}\geq \frac{m-l}{m}x_0 \Big)$. For some specific $t>0$,
\begin{align}\label{ap:14}
&\mathbbm{P}\Big(\bm{A}+\bm{B}\geq \frac{m-l}{m}x_0 \Big) = \mathbbm{P}\Big(\bm{A_1}+\bm{A_2}+\bm{A_3}+\bm{A_4}+\bm{B_1}+\bm{B_2}\geq \frac{m-l}{m}x_0 \Big)\nonumber\\
\leq &\mathbbm{P}\Big(\bm{A_1}+\bm{B_1}\geq t \Big)+\mathbbm{P}\Big(\bm{A_3}+\bm{B_2}\geq \frac{m-l}{m}x_0  - t - \bm{A_2}-\bm{A_4}\Big).
\end{align}
We then transfer $\bm{A_3}+\bm{B_2}$ into a quadratic form. For each community $q, q = 1,\cdots,K$ define the transformations:
$$
f_q^+(x) = \sqrt{\Big[z^*_{iq}\{Y_{is}^t \log\frac{\mu_{qa}}{\mu_{qb}} + (1-Y_{is}^t) \log\frac{1-\mu_{qa}}{1-\mu_{qb}}\}\Big]_{+}},$$
$$f_q^-(x) = \sqrt{\Big[z^*_{iq}\{Y_{is}^t \log\frac{\mu_{qa}}{\mu_{qb}} + (1-Y_{is}^t) \log\frac{1-\mu_{qa}}{1-\mu_{qb}}\}\Big]_{-}},$$
$$X_t^+ = \{f_1^+(Y^t_{1s}),\cdots,f_1^+(Y^t_{ns}),f_2^+(Y^t_{1s}),\cdots, f_2^+(Y^t_{Ns}),\cdots,f_K^+(Y^t_{1s}),\cdots,f_K^+(Y^t_{Ns})\},$$
$$X_t^- = \{f_1^-(Y^t_{1s}),\cdots,f_1^-(Y^t_{Ns}),f_2^-(Y^t_{1s}),\cdots, f_2^-(Y^t_{Ns}),\cdots,f_K^-(Y^t_{1s}),\cdots,f_K^-(Y^t_{Ns})\}.
$$ 
Notice that the total number of non-zero terms in $X_t^+$ or $X_t^-$ is $N$. 
We define the node sets 
\begin{align*}
\tilde{\xi}_a = \{(i_1,i_2,i_3)|z^*_{i_1a}z^*_{i_2a}z^*_{i_3a} = 1\} \;\;\;
\tilde{\xi}_b = \{(i_1,i_2,i_3)|z^*_{i_1b}z^*_{i_2b}z^*_{i_3b} = 1\}.
\end{align*}
Note $\#|\tilde{\xi}_a| = o(N_a^3)$ and $\#|\tilde{\xi}_b| = o(N_b^3)$ where $N_a$ and $N_b$ are number of node in community $a$ and $b$. We further define augmented edges vectors:
\begin{align*}
\bar{X}_t^+ = \Bigg(X_t^+, \underbrace{\Big(\frac{\rho C_A}{4}\sqrt{\{\hat{Y}_{i_1s}^{t,a}\hat{Y}_{i_2i_3}^{t,a}\}_{+}}\Big)_{1\times \#|\tilde{\xi}_a|}}_{\substack{(i_1,i_2,i_3)\in \tilde{\xi}_a}}, \underbrace{\Big(\frac{\rho C_A}{4}\sqrt{\{-\hat{Y}_{i_1s}^{t,b}\hat{Y}_{i_2i_3}^{t,b}\}_{+}}\Big)_{1\times \#|\tilde{\xi}_b|}}_{\substack{(i_1,i_2,i_3)\in \tilde{\xi}_b}}  \Bigg),\\
\bar{X}_t^- = \Bigg(X_t^-, \underbrace{\Big(\frac{\rho C_A}{4}\sqrt{\{\hat{Y}_{i_1s}^{t,a}\hat{Y}_{i_2i_3}^{t,a}\}_{-}}\Big)_{1\times \#|\tilde{\xi}_a|}}_{\substack{(i_1,i_2,i_3)\in \tilde{\xi}_a}}, \underbrace{\Big(\frac{\rho C_A}{4}\sqrt{\{-\hat{Y}_{i_1s}^{t,b}\hat{Y}_{i_2i_3}^{t,b}\}_{-}}\Big)_{1\times \#|\tilde{\xi}_b|}}_{\substack{(i_1,i_2,i_3)\in \tilde{\xi}_b}}  \Bigg).
\end{align*}
Denote the covariance of $\bar{X}_t^+$ and $\bar{X}_t^-$ as $\bar{\Sigma}_1$ and $\bar{\Sigma}_2$. Note that each element in $\bar{X}_t^+$ or $\bar{X}_t^-$ is a bounded binary random variable. Similarly, $\bar{X}_t^+$ and $\bar{X}_t^-$ are subgaussian vectors.
Therefore,
\begin{align*}
\bm{A_3}+\bm{B_2} &= \frac{1}{M}\sum_{t=1}^M \big(\langle \bar{X}_t^+, \bar{X}_t^+ \rangle - \langle \bar{X}_t^-, \bar{X}_t^- \rangle \big) = \frac{1}{M}(\bar{Q}_1 - \bar{Q}_2),\\
E(\bm{A_3}+\bm{B_2}) &= \frac{1}{M}(E\bar{Q}_1 - E\bar{Q}_2).
\end{align*}
Denote $s = \frac{m-l}{m}x_0 - t - \bm{A_2}-\bm{A_4} - E(\bm{A_3}+\bm{B_2})$, we estimate $E(\bm{A_3}+\bm{B_2})$, $\bm{A_2}$ and $\bm{A_4}$ in the following. 
Given $z^*_s = b$ and the result in (\ref{ap:3}), we have for some constant $c>0$ and $q = 1,\cdots, K$:
\begin{align*}
E\Big[\{Y_{is}^t \log\frac{\mu_{qa}}{\mu_{qb}} + (1-Y_{is}^t) \log\frac{1-\mu_{qa}}{1-\mu_{qb}}\}\Big] =
\mu_{qb}\log\frac{\mu_{qa}}{\mu_{qb}} + (1-\mu_{qb}) \log\frac{1-\mu_{qa}}{1-\mu_{qb}} < -c < 0.
\end{align*}
Then 
\begin{align*}
E\bm{A_3} = \frac{1}{M}\sum_{t=1}^M\sum_{q=1}^K\sum_{i\neq s}^N\Big[z^*_{iq}\{\mu_{qb} \log\frac{\mu_{qa}}{\mu_{qb}} + (1-\mu_{qb}) \log\frac{1-\mu_{qa}}{1-\mu_{qb}}\}\Big] < -c(N-1).
\end{align*}
Given edges from different communities are independent and correlation density $\lambda$, there exists a constant $C>0$ such that
\begin{align*}
E\bm{B_2} = &\frac{\rho C_A}{4}\Big[\langle \alpha_a\otimes vec(\alpha_a^T\alpha_a), E\{\hat{Y}^{t,a}_{\cdot s}\otimes vec(\hat{\bm{Y}}^{t,a})\} \rangle \!-\! \langle   \alpha_b\otimes vec(\alpha_b^T\alpha_b), E\{\hat{Y}^{t,b}_{\cdot i}\otimes vec(\hat{\bm{Y}}^{t,b})\}\rangle \Big]\\
 = & -\frac{\rho C_A}{4} \langle   \alpha_b\otimes vec(\alpha_b^T\alpha_b), E\{\hat{Y}^{t,b}_{\cdot i}\otimes vec(\hat{\bm{Y}}^{t,b})\}\rangle \leq -C\lambda N_b^3.
\end{align*} 
Therefore, $-E(\bm{A_3}+\bm{B_2})\geq c(N-1)+C\lambda (\kappa_1N)^3 \geq c'(N+\lambda N^3)$ for some positive constant $c'$.
Based on condition C1 that $\mu_{ql}, q,l = 1,\cdots, K$ are bounded, it can be shown that $|EY_{is}^t \log\frac{\mu_{qa}}{\mu_{qb}} + (1-EY_{is}^t) \log\frac{1-\mu_{qa}}{1-\mu_{qb}}|$ is bounded then $|\bm{A_2}|= \mathcal{O}_N(N)$.\\

From condition C5, we have 
\begin{align*}
\log\frac{\gamma_1}{\gamma_2}\leq \log\frac{\mu_{qa}\hat{\mu}_{qb}}{\hat{\mu}_{qa}\mu_{qb}}\leq \log\frac{\gamma_2}{\gamma_1} \;\;\; \text{and}\;\;\;\log\frac{1-\gamma_2}{1-\gamma_1}\leq\log\frac{(1-\mu_{qa})(1-\hat{\mu}_{qb})}{(1-\hat{\mu}_{qa})(1-\hat{\mu}_{qb})}\leq \log\frac{1-\gamma_1}{1-\gamma_2}
\end{align*}
Define $\gamma = \max\{-\log\frac{\gamma_1}{\gamma_2}, \frac{\gamma_2}{\gamma_1},  -\frac{1-\gamma_2}{1-\gamma_1}, \frac{1-\gamma_1}{1-\gamma_2}\}$. Then we have
\begin{align*}
|\bm{A_4}| = &|\frac{1}{M}\sum_{t=1}^M\sum_{q=1}^K\sum_{i\neq s}^N\Big[\alpha_{iq}\{Y_{is}^t \log\frac{\mu_{qa}\hat{\mu}_{qb}}{\hat{\mu}_{qa}\mu_{qb}} + (1-Y_{is}^t) \log\frac{(1-\mu_{qa})(1-\hat{\mu}_{qb})}{(1-\hat{\mu}_{qa})(1-\hat{\mu}_{qb})}\}\Big]|\\
\leq & \gamma|\sum_{q=1}^K\sum_{i\neq s}^N\alpha_{iq}| \leq \gamma N
\end{align*}
Therefore we have $|\bm{A_2}+\bm{A_4} | = \mathcal{O}_N(N)$. We choose $t = -\frac{E(\bm{A_3}+\bm{B_2})}{2}$ and $x_0 = -\alpha'MN$ where $\alpha'>0$. As the function of node size $N$, $M$ and $\lambda$ are constrained in the range $M\leq o(N^{2-\frac{\eta}{2}})$ and $\lambda N^{\frac{\eta}{2}} >1$, where $\eta$ is defined in condition C4. Then $\frac{m-l}{m}x_0 = o_N(E(\bm{A_3}+\bm{B_2}))$. Therefore, $E(\bm{A_3}+\bm{B_2})$ is dominant term in $s$ such that $s \geq -C'\lambda N^3$ where $C'>0$ is a constant. Follow a similar discussion in (\ref{ap:4}) and condition C3, we have the upper bound for $\|\bar{\Sigma}_1\|_{op}$:
\begin{align*}
\|\bar{\Sigma}_1\|_{op}\leq c_0(1+c_1\lambda N^2).
\end{align*}
In addition, from $\#|X_t^+| = N$, $\#|\bar{\xi}_a| = o(N_a^3)$, $\#|\bar{\xi}_b| = o(N_b^3)$ and condition C3, we have the upper bound for $\|\bar{\Sigma}_1\|_{F}^2$:
\begin{align*}
\|\bar{\Sigma}_1\|_{F}^2 \leq C_1N(1+c\lambda N^2)+C_2N^3(1+c^*\lambda N^2),
\end{align*}
where $C_1, C_2, c, c^*$ are constants.
Then we estimate the upper bound for the second term in (\ref{ap:14}):
\begin{align*}
&\mathbbm{P}\Big(\bm{A_3}+\bm{B_2}\geq \frac{m-l}{m}x_0  - t - \bm{A_2}- \bm{A_4}\Big) = \mathbbm{P}\Big\{(\bar{Q}_1-E\bar{Q}_1)-(\bar{Q}_2-E\bar{Q}_2)> Ms \Big\} \\
\leq & \frac{1}{2} \mathbbm{P}\Big\{ |\bar{Q}_1-E\bar{Q}_1| > \frac{Ms}{2}\Big\} + \frac{1}{2} \mathbbm{P}\Big\{ |\bar{Q}_2-E\bar{Q}_2| > \frac{Ms}{2}\Big\}.
\end{align*}
According to the generalized Hanson-Wright inequality in (\cite{chen2018hanson}):
\begin{align}
&\frac{1}{2} \mathbbm{P}\Big\{ |\bar{Q}_1-E\bar{Q}_1| > s \Big\} \leq \exp\Big\{-C\min\big(\frac{s^2M^2}{\bar{L}_1^4\|\bar{\Sigma}_1\|_{F}^2\|A\|_{F}^2},\frac{sM}{\bar{L}_1^2\|\bar{\Sigma}_1\|_{op}\|A\|_{op}}\big)\Big\},
\end{align}
where $A = \bm{I}_{M\times M}$ and $\bar{L}_1$ is subgaussian norm of $\bar{X}_t^+$. Notice that
\begin{align*}
\frac{s^2M^2}{\bar{L}_1^4\|\bar{\Sigma}_1\|_{F}^2\|A\|_{F}^2} \geq \frac{(C'\lambda N^3)^2M^2}{\bar{L}_1^4 \{C_1N(1+c\lambda N^2)+C_2N^3(1+c^*\lambda N^2)\}M}, 
\frac{sM}{\bar{L}_1^2\|\bar{\Sigma}_1\|_{op}\|A\|_{op}} \geq \frac{C'\lambda N^3M}{\bar{L}_1^2c_0(1+c_1\lambda N^2)}.
\end{align*}
Given $\lambda N^{\frac{\eta}{2}} > 1$, we have for some constant $C^*>0$ 
\begin{align*}
C\min\big(\frac{s^2M^2}{\bar{L}_1^4\|\bar{\Sigma}_1\|_{F}^2\|A\|_{F}^2},\frac{sM}{\bar{L}_1^2\|\bar{\Sigma}_1\|_{op}\|A\|_{op}}\big) \geq C^*\lambda MN.
\end{align*}
The upper bound for $\mathbbm{P}\Big\{ |\bar{Q}_2-E\bar{Q}_2| > \frac{Ms}{2}\Big\}$ can be similarly obtained. Therefore, 
\begin{align*}
\mathbbm{P}\Big(\bm{A_3}+\bm{B_2}\geq \frac{m-l}{m}x_0  - t - \bm{A_2}\Big) \leq \exp(-C'\lambda MN).
\end{align*}
Next, we estimate the term $\mathbbm{P}\Big( \bm{A_1}+\bm{B_1}\geq t\Big)$. Notice 
\begin{align*}
E(\bm{A_1}) = E\Big[\frac{1}{M}\sum_{t=1}^M\sum_{q=1}^K\sum_{i\neq s}^N(\alpha_{iq} -z_{iq}^*)\{Y_{is}^t - E(Y_{is}^t)\}(\log\frac{\mu_{qa}}{\mu_{qb}}-\log\frac{1-\mu_{qa}}{1-\mu_{qb}})\Big] = 0,\\
E(\bm{B_1}) = \frac{\rho C_A}{4M}\sum_{t=1}^M\big[\langle \alpha_a\otimes vec(\alpha^T_a\alpha_a) - z_a^*\otimes vec(z_a^{*T}z_a^*), E\{\hat{Y}^{t,a}_{\cdot s}\otimes vec(\hat{\bm{Y}}^{t,a})\}\rangle-\nonumber \\\langle \alpha_b\otimes vec(\alpha^T_b\alpha_b) - z_b^*\otimes vec(z_b^{*T}z_b^*),E\{\hat{Y}^{t,b}_{\cdot s}\otimes vec(\hat{\bm{Y}}^{t,b})\}\rangle\big].
\end{align*}
Given condition C4 such that $\|\bm{\alpha}-z^*\|_1 = cN^{1-\eta}, 0<\eta<1$, 
\begin{align*}
\bm{B}_1 = \frac{\rho C_A}{4M}\sum_{t=1}^M&\big\{\langle \alpha_a\otimes vec(\alpha_a^T\alpha_a) - z_a^*\otimes
vec(z_a^{*T}z_a^*), \hat{Y}^a_{\cdot s}\otimes vec(\hat{\bm{Y}}^a)\rangle-\\
&\langle \alpha_b\otimes vec(\alpha^T_b\alpha_b) - z_b^*\otimes vec(z_b^{*T}z_b^*), \hat{Y}^b_{\cdot s}\otimes vec(\hat{\bm{Y}}^b)\rangle\big\}.
\end{align*} 
Notice that for any community $a = 1,\cdots, K$,
\begin{align*}
\|(vec(\alpha_a^T\alpha_a) -vec(z_a^{*T}z_a^*))\|_2&\leq \|\alpha_a\otimes(\alpha_a-z_a^*)\|_2+\|(\alpha_a-z_a^*)\otimes z_a^*\|_2\\
& \leq \|\alpha_a\|_2\|(\alpha_a-z_a^*)\|_2+\|(\alpha_a-Z_a^*)\|_2\| z_a^*\|_2,\\
\|E(\hat{Y}^{t,a}_{\cdot s})\|_2&\leq \sqrt{\frac{N}{\hat{\mu}_{aa}(1-\hat{\mu}_{aa})}},\;\;\;
\|E(\hat{\bm{Y}}^{t,a})\|_2\leq \sqrt{\frac{N^2}{\hat{\mu}_{aa}(1-\hat{\mu}_{aa})}}.
\end{align*}
Therefore, we have
\begin{align*}
\;\;\;\;\;\;&\langle \alpha_a\otimes vec(\alpha_a^T\alpha_a) - z_a^*\otimes vec(z_a^{*T}z_a^*), E\{\hat{Y}^{t,a}_{\cdot s}\otimes vec(\hat{\bm{Y}}^{t,a})\}\rangle  \\
\leq& \| \alpha_a\otimes vec(\alpha_a^T\alpha_a) - z_a^*\otimes vec(z_a^{*T}z_a^*) \|_2 \|E\{\hat{Y}^{t,a}_{\cdot s}\otimes vec(\hat{\bm{Y}}^{t,a})\}\|_2 \\
\leq &\big( \| \alpha_a\otimes vec(\alpha_a^T\alpha_a) -vec(z_a^{*T}z_a^*))\|_2+ \|(\alpha_a-z_a^*)\otimes vec(z_a^{*T}z_a^*)\|_2\big)\|E\{\hat{Y}^{t,a}_{\cdot s}\otimes vec(\hat{\bm{Y}}^{t,a})\}\|_2\\
\leq &\|\alpha_a - z_a^*\|_2\!\cdot\!\big( \|\alpha_a\|_2^2+\|z_a^*\|_2^2+\|\alpha_a\|_2\|z_a^*\|_2\big)\!\cdot\!\|E(\hat{Y}^{t,a}_{\cdot s})\|_2\cdot\!\|E(\hat{\bm{Y}}^{t,a})\|_2
\leq  \frac{3N*N^{3/2}}{\hat{\mu}_{aa}(1-\hat{\mu}_{aa})}\|\alpha_a - z_a^*\|_2.
\end{align*}
Since $\|\alpha_a - z_a^*\|_2 = \sqrt{\|\alpha_a - z_a^*\|_2^2} \leq \sqrt{\|\bm{\alpha} - z^*\|_1}$ for any $a = 1,\cdots, K$, then for some constant $C>0$,
\begin{align*}
|E(\bm{B}_1)| \leq CN^{3-\frac{\eta}{2}}.
\end{align*}
We define edge vectors $\tilde{Y}_t, t = 1,\cdots,M$ and membership vector $\bm{\theta}_{a,b}$ as:
\begin{align*}
\tilde{Y}_t = \big\{\underbrace{Y^t_{\cdot s}-E(Y^t_{\cdot s}),\cdots, Y^t_{\cdot s}-E(Y^t_{\cdot s})}_{NK}, \hat{Y}^{t,a}_{\cdot s}\otimes vec(\hat{\bm{Y}}^{t,a}),\hat{Y}^{t,b}_{\cdot s}\otimes vec(\hat{\bm{Y}}^{t,b})\big\},\\
\bm{\theta}_{a,b} = \Big[\underbrace{(\alpha_{iq} -z_{iq}^*)(\log\frac{\mu_{qa}}{\mu_{qb}}-\log\frac{1-\mu_{qa}}{1-\mu_{qb}})}_{i=1\cdots, N},\cdots, \underbrace{(\alpha_{iK} -z_{iK}^*)(\log\frac{\mu_{Ka}}{\mu_{Kb}}-\log\frac{1-\mu_{Ka}}{1-\mu_{Kb}})}_{i=1\cdots, N},\\ \frac{\rho C_A}{4}\{\alpha_a\otimes vec(\alpha^T_a\alpha_a) - z_a^*\otimes vec(z_a^{*T}z_a^*)\},\frac{\rho C_A}{4}\{\alpha_b\otimes vec(\alpha^T_b\alpha_b) - z_b^*\otimes vec(z_b^{*T}z_b^*)\}\Big].
\end{align*}

Notice for $a,b = 1,\cdots, K$, we have
\begin{align*}
\|\bm{\theta}_{a,b}\|_2^2 &\leq \mu_2\|\bm{\alpha}-z^*\|_2^2 \!+ \!\|\alpha_a\otimes vec(\alpha^T_a\alpha_a) \!-\! z_a^*\otimes vec(z_a^{*T}z_a^*)\|_2^2 \!+\! \|\alpha_b\otimes vec(\alpha^T_b\alpha_b) \!-\! z_b^*\otimes vec(z_b^{*T}z_b^*)\|_2^2\\
&\leq \mu_2\|\bm{\alpha}-z^*\|_1 + C_1N^2(\|\alpha_a - z_a^*\|_1+\|\alpha_b - z_b^*\|_1),
\end{align*}
where $\mu_2: = \max\{(\log\frac{\mu_{qa}}{\mu_{qb}}-\log\frac{1-\mu_{qa}}{1-\mu_{qb}})\},\; q = 1,\cdots,K$ and $C_1>0$ is a constant.
Then we can transform $\var(\bm{A_1}+\bm{B_1})$ into
\begin{align*}
\var(\bm{A_1}+\bm{B_1}) = \frac{1}{M}\sum_{t=1}^M\var(\bm{\theta}_{a,b}\tilde{Y}_t) = \frac{1}{M}\sum_{t=1}^M\bm{\theta}_{a,b}^T Cov(\tilde{Y}_t, \tilde{Y}_t)\bm{\theta}_{a,b} \leq \frac{1}{M}\|Cov(\tilde{Y}_t, \tilde{Y}_t)\|_{op}\|\bm{\theta}_{a,b}\|_2^2.
\end{align*}
From the condition C3 and same discussion in (\ref{ap:4}), we have for some constant $C>0$ and $c>0$:
\begin{align*}
\|Cov(\tilde{Y}_t, \tilde{Y}_t)\|_{op} \leq C(1+c\lambda N^2).
\end{align*}
Given $\frac{1}{\lambda} = o(N^{\frac{\eta}{2}})$, we have $E(\bm{A_1}+\bm{B_1}) = o_N(E(\bm{A_3}+\bm{B_2}))$ then the $E(\bm{A_3}+\bm{B_2})$ is dominating in the term ${\{t - E(\bm{A_1}+\bm{B_1})\}^2}$. Based on the Markov inequality, for some constant $C_2>0$
\begin{align*}
\mathbbm{P}\Big( \bm{A_1}+\bm{B_1}\geq t\Big)\leq& \frac{\var(\bm{A_1}+\bm{B_1})}{\{t - E(\bm{A_1}+\bm{B_1})\}^2}\leq \frac{\|Cov(\tilde{Y}_t, \tilde{Y}_t)\|_{op}\|\bm{\theta}_{a,b}\|_2^2}{M\{ c'(N+\lambda N^3)\}^2}\\
\leq & \frac{C(1+c\lambda N^2)\{\mu_2\|\bm{\alpha}-z^*\|_1 + C_1N^2(\|\alpha_a - z_a^*\|_1+\|\alpha_b - z_b^*\|_1)\}}{( c'(N+\lambda N^3))^2M}\\
\leq & \frac{2Cc\{\mu_2\|\bm{\alpha}-z^*\|_1 + C_1N^2(\|\alpha_a - z_a^*\|_1+\|\alpha_b - z_b^*\|_1)\}}{c'^2(1+\sqrt{\lambda}N^2)^2M}\\
\leq & C_2\frac{N^{\eta/4}(\|\alpha_a - z_a^*\|_1+\|\alpha_b - z_b^*\|_1)}{(1+\lambda  N^{2+\frac{\eta}{4}})M}.
\end{align*}
Combined upper bound of $\mathbbm{P}\Big( \bm{A_1}+\bm{B_1}\geq t\Big)$ and $\mathbbm{P}\Big( \bm{A_3}+\bm{B_2}\geq s\Big)$ with (\ref{ap:13}), there exists positive constant $c_1>0, c_2>0, c_3>0$ such that:
\begin{align*}
&E\|\bm{\alpha}^{s+1} - \bm{z^*}\|_1 \leq 2\bm{w}NK\exp(-\alpha' MN) + 2\bm{w}\sum_{l=0}^{m-1}\sum_{a=1}^K\sum_{b\neq a}\sum_{i:z_i^*= b}\frac{1-exp(\alpha' MN)}{m}E(\bm{L}_2)\\
\leq & 2\bm{w}KN\exp(-\alpha' MN)\! +\! 2\bm{w}mKN\exp(-C'\lambda MN)\!+\! 2\bm{w}mKNC_2\frac{N^{\eta/4}(\|\alpha_a \!-\! z_a^*\|_1\!+\!\|\alpha_b\ !-\! z_b^*\|_1)}{(1+\lambda  N^{2+\frac{\eta}{4}})M}\\
\leq & c_1NK\exp(-c_2(1+\lambda)MN)+ \frac{c_3N^{1+\frac{\eta}{4}}\|\bm{\alpha}^{s} - \bm{z^*}\|_1}{(1+\lambda  N^{2+\frac{\eta}{4}})M}.
\end{align*}

\begin{supplement}
\sdescription{\textbf{Supplement to "Community detection with dependent connectivity"}. Due  to  space  constraints, we relegate proofs of Lemma 1 and Lemma 2 to the supplement.}
\end{supplement}

%

\bibliography{manuscript_Annal}

\end{document}